%% file: main.tex
\documentclass[%
reprint,
superscriptaddress,
amsmath,
amssymb,
aps,
pra,
floatfix,
]{revtex4-2}
\usepackage{graphicx} 
\usepackage{dcolumn} 
\usepackage{bm}
\usepackage{hyperref}
\usepackage{xcolor}

\newcommand{\up}{\uparrow}
\newcommand{\down}{\downarrow}
\newcommand{\spsm}{\ensuremath{ \sigma_{\scriptscriptstyle +} \!-\! \sigma_{\scriptscriptstyle -} }  }

\newcommand{\del}[1]{}

\newcommand{\affchem}{Department of Chemistry, Purdue University, West Lafayette, Indiana, 47907, USA}
\newcommand{\affphys}{Department of Physics and Astronomy, Purdue University, West Lafayette, Indiana, 47907, USA}

\begin{document}

\preprint{APS/123-QED}

\title{A Generalized Theory for Optical Cooling of a Trapped Atom with Spin}

\author{Saumitra S. Phatak}
\affiliation{\affphys}
\author{Karl N. Blodgett}
\affiliation{\affchem}
\author{David Peana}
\affiliation{\affchem}
\author{Meng Raymond Chen}
\affiliation{\affchem}
\author{Jonathan D. Hood}
\affiliation{\affphys, \affchem}
\email{hoodjd@purdue.edu}
\date{\today}

\begin{abstract}
Cooling atoms to the ground state of optical tweezers is becoming increasingly important for high-fidelity imaging, cooling, and molecular assembly.  While extensive theoretical work has been conducted on cooling in free space, fewer studies have focused on cooling in bound states. In this work, we present a unified formalism for optical cooling mechanisms in neutral atom tweezers, including resolved and unresolved sideband cooling with different trapping potentials, polarization gradient cooling, gray molasses cooling, $\Lambda$-enhanced gray molasses cooling, and Raman sideband cooling. We perform simulations  {with hyperfine levels} and demonstrate good agreement with a simplified spin model. We derive and discuss the fundamental limits of each cooling mechanism and propose new strategies for achieving ground state cooling in optical tweezers. Our findings provide valuable insights into optimizing cooling schemes for neutral atoms in optical tweezers, paving the way for minimizing thermal decoherence in Rydberg and molecular gates and improving efficiencies of molecular assembly. 
\end{abstract}

\maketitle

%%%%%%%%%%%%%%%%%%%%%%%%%%%%%%%%%%%%%%%%%%%%%%%%%%%%%%%%%%%%%%%%%%
\section{Introduction}  \label{sec:intro}
Laser cooling of bound neutral atoms has recently gained renewed significance in experiments involving tweezer arrays of atoms and molecules~\cite{kaufman2021quantum}. In systems of Rydberg atoms~\cite{bluvstein2024logical} or molecules~\cite{anderegg2019optical, bao2023dipolar, Ruttley2024Enhanced} with dipolar coupling, thermal motion limits the coherence of interactions. Molecular assembly~\cite{zhang2022optical, molony2014creation} and few-body physics~\cite{spar2022realization, holten2021observation} studies require high-fidelity ground state preparation of multiple atoms.

\begin{figure*}[t]
   \centering
    \includegraphics[width=0.99\linewidth]{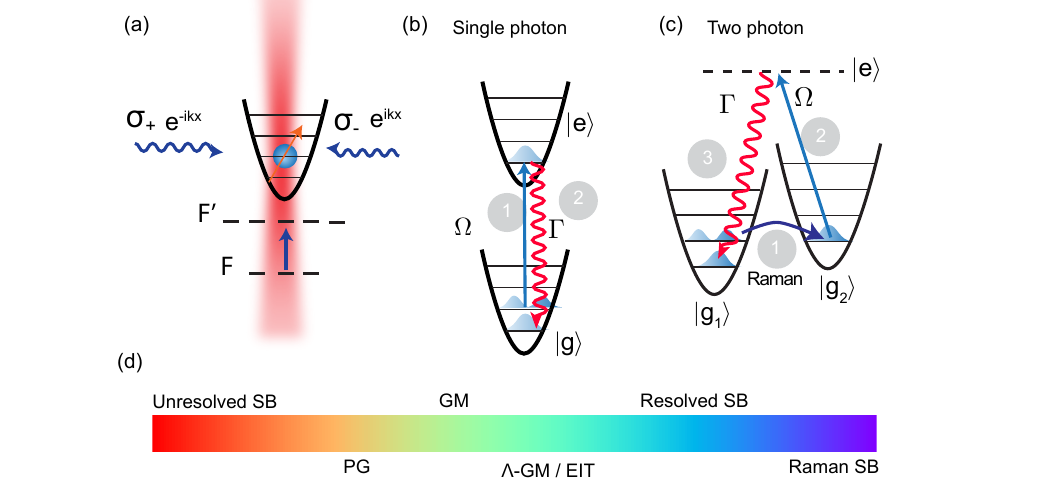}
    \caption{  \textbf{(a)} Laser cooling of a trapped atom with ground and excited state spin $F$ and $F'$ in a changing polarization from counter-propagating beams.  \textbf{(b)} Single photon cooling schemes such as resolved and unresolved sideband cooling drive population from the ground state to the excited state with one reduced motional quanta $n-1$.  In the Lamb-Dicke regime, the recoil heating is smaller than the trapping frequency and the population mostly returns to $n-1$, resulting in net cooling. \textbf{(c)}  Two-photon cooling schemes like polarization gradient (PG), gray molasses (GM), and $\Lambda$-enhanced gray molasses ($\Lambda$-GM) drive $n \rightarrow n-1$ between two ground state spins with a two-photon Raman transition.  \textbf{(d)} Temperature scale for cooling schemes.  A wide variety of cooling schemes has been used to trap, cool, and image tightly trapped atoms.   \del{Take out n and n-1, add Raman to (c)}  \del{Can we reference a,b,c,d}
    }
    \label{fig:intro}
\end{figure*}

A wide range of optical cooling techniques have been developed for trapped atoms, illustrated in Fig.~\ref{fig:intro}. Single-photon cooling schemes, such as sideband cooling, have been used for atoms with narrow lines for high-fidelity imaging and ground state preparation~\cite{cooper2018alkaline,saskin2019narrow,jenkins2022ytterbium}. Two-photon schemes, including polarization gradient (PG)~\cite{dalibard1989laser, tuchendler2008energy, lester2015rapid, endres2016atom}, 
gray molasses (GM) and $\Lambda$-enhanced gray molasses ($\Lambda$-GM)~\cite{salomon2013gray, grier2013lambda, Yue2017Polarization, Brown2019Gray, blodgett2023imaging}, and Raman-sideband cooling (RSB)~\cite{kerman2000beyond, Popp2006Ground, kaufman2012cooling, thompson2013coherence, yu2018motional}, have been successfully applied for imaging and cooling of atoms without narrow lines.  While extensive theoretical work has been developed for sideband cooling~\cite{monroe1995resolved, schliesser2008resolved, peik1999sideband}, many of the other cooling schemes for bound atoms are still interpreted through their free-space pictures, for example in Sisyphus cooling where an atom moves through a changing polarization~\cite{dalibard1989laser}. However, this picture breaks down for a tightly trapped atom in the Lamb-Dicke regime, where the atom wavefunction is smaller than the wavelength.

In this work, we develop a generalized optical cooling model for a bound atom with spin in a single or counter-propagating beam configuration with different polarizations. We validate these models using master equation simulations that include the ground and excited spins and the harmonic oscillator states.
Our analysis reveals a consistent picture across various cooling schemes, shown in Fig.~\ref{fig:intro}.  Cooling occurs when $|n \rangle$ and $|n-1 \rangle $ are Raman coupled for different spin states and brought into resonance using a combination of light shifts (PG, GM, $\Lambda$-GM) and two-photon detunings (RSB). 

Sideband cooling has been extensively studied in previous works~\cite{stenholm1986thesemiclassical, leibfried2003quantum}, but  {studies for} cooling \del{of} a bound atom with spin has been much more limited.   
Previous  {theoretical} work on cooling bound atoms with spin has primarily focused on specific cases, such as $J\!\!=\!\!1/2$ to $J'\!\!=\!\!3/2$ transitions in counter-propagating orthogonal linear light~\cite{cirac1993laser} or in a linear light standing wave~\cite{wineland1992sisyphus}. We develop a new formalism for arbitrary spin and polarization, with a particular emphasis on the experimentally relevant $\sigma_+ \! - \! \sigma_-$ configuration. Notably, while these previous theoretical studies found a fundamental limit of $\langle n \rangle \approx 1$, we demonstrate that schemes like GM and $\Lambda$-GM can be used for high ground state preparation when considering different polarization and spin configurations.

A recent experimental result achieved a record-high imaging fidelity of 99.96\% for lithium using $\Lambda$-enhanced GM cooling with significant ground state preparation, despite lithium having the largest recoil heating~\cite{blodgett2023imaging}. This raises the question of whether spin cooling techniques can be further optimized for ground state preparation. The unified model also allows us to explore novel cooling schemes that transcend the conventions of any single conventional cooling technique.  

The paper is structured as follows. Section~\ref{sec:theory} develops the general master equation for spin cooling and derives effective ground state spin operators by adiabatically eliminating the excited state. Section~\ref{sec:sideband} applies this formalism to sideband cooling, including cases with different excited state trapping frequencies. Section~\ref{sec:spin} develops a formalism for spin cooling and applies it to PG, GM, $\Lambda$-GM, and RSB cooling. Each section compares exact simulations to simplified models and contrasts various cooling techniques, with the goal of creating a unified picture that will lead to more coherent interactions and higher fidelity ground state preparation.

%%%%%%%%%%%%%%%%%%%%%%%%%%%%%%%%%%%%%%%%%%%%%%%%%%%%%%%%%%%%%%%%%%
\section{Theory: Laser cooling in a harmonic potential} \label{sec:theory}
\subsection{Master equation for laser cooling} \label{subsec:master}
In this section, we develop a master equation for an atom confined in a tight harmonic trap and interacting with a semi-classical electric field. In a semi-classical approximation, we take the expectation of the electric field but keep the atom position  $\hat{x}$ as a quantum operator~\cite{phillips1992laser, molmer1996monte}.  % see gradiner quantm noise pg 389 for more details.  % phillips 1992 is cite Dalibard, Raimin, Zinn-Jutin Fnumdanetal system Les Houche 1990, pub 1992,   Kazantseu, Mechanical   % gardiner2004quantum
The electric field for a single laser frequency $\omega_L$ is:
\begin{equation}
\bm{E}(\hat{x} , t) = E_0 \, {\boldsymbol{\mathcal{\epsilon}}}( \hat{x} ) \,  e^{- i \omega_L t}  + h.c.  
\end{equation}  
$E_0$ is a real field amplitude and the last term is the hermitian conjugate. The complex polarization and phase are contained within $\bm{\epsilon}(\hat{x})$ and can describe a traveling wave or a standing wave with\del{various}  {arbitrary} polarization configurations. 

The atom has ground hyperfine spin states denoted as \(|F_g m_g\rangle\) and excited hyperfine spin states denoted as \(|F_e m_e\rangle\). The ground and excited state projection operators \(P_e\) are
\begin{gather}
    P_e = \sum_{m_e} |F_e m_e\rangle \langle F_e m_e| \nonumber \\
    P_g = \sum_{m_g} |F_g m_g\rangle \langle F_g m_g|.
\end{gather}
For an optically trapped atom, the ground and excited states can generally experience different potentials due to their differing polarizabilities. This difference will be addressed later, but for now, we will assume that the ground and excited states share the same harmonic trapping frequency \(\nu\).
The total Hamiltonian for the bound atom interacting with the electric field in the rotating frame of the laser is~\cite{phillips1992laser, molmer1996monte}  
\begin{gather} \label{eq:H}
H = \nu \hat{a}^\dag \hat{a}  +   \left( \Delta - i \frac{\Gamma}{2} \right) P_e  - \frac{\Omega}{2}  \left( \boldsymbol{\mathcal{\epsilon}}(\hat{x} ) \cdot \boldsymbol{\hat{D}}^\dag + h.c. \right).
\end{gather}
 $\Delta = \omega_A - \omega_L$ is the detuning of the atom from the laser.
The Rabi frequency $\Omega$ is defined in terms of the reduced matrix element times the electric field amplitude 
\begin{equation}
    \Omega = \langle J_e || d || J_g \rangle  E_0/\hbar.   
\end{equation} 
The atomic operator $\hat{\bm{D}}^\dag$ is a raising operator for all the hyperfine levels in the ground state  {to the excited state}, and $\hat{\bm{D}}$ is the corresponding lowering operator.  They are defined in terms of the dipole operator $\boldsymbol{\hat{d}}$ normalized by the reduced matrix element,
\begin{equation} \label{eq:dipoleoperator}
    \hat{\boldsymbol{D}} =  \frac{P_{g} \boldsymbol{\hat{d}} P_{e} }{\langle J_e || d ||  J_g\rangle  },  \quad  \hat{\boldsymbol{D}}^\dag =  \frac{P_{e} \boldsymbol{\hat{d}} P_{g} }{\langle J_e || d ||  J_g\rangle  }.  
\end{equation}

The Hamiltonian in Eq.~\ref{eq:H} also contains a non-Hermitian imaginary decay rate $\Gamma$ of the excited state due to spontaneous emission. In the Wigner-Weisskopf model, the excited state decays due to spontaneous emission without repopulating the ground state. This leads to a reduction in the wavefunction normalization.  The master equation for the density matrix contains refeeding terms with collapse operators $L_i$ that repopulate the ground state and preserve the density matrix normalization:
\begin{equation} \label{eq:master}
    \frac{d \hat{\rho}}{dt} = -i ( H \hat{\rho} - \hat{\rho} H^\dag ) +  \sum_{i} {L}_i \hat{\rho} {L}^\dag_i.
\end{equation}
The collapse operators contain both the lowering of the excited state as well as the momentum recoil for emitting a photon:
\begin{equation}
    {L}_{\hat{\bm{n}},q} = \sqrt{\Gamma} \, \hat{R}_{\hat{\bm{n}},q} \hat{D}_q.
\end{equation}
The collapse operators have two effects when operating on the density matrix. First, they lower the atom back to the ground state. $\hat{D}_q$ is the lowering operator in the spherical vector basis $\hat{D}_q = \hat{\bm{e}}^*_q \cdot \hat{\bm{D}}$, where $\hat{\bm{e}}_q$ is a spherical basis vector. Second, the recoil operator $\hat{R}_{n,q}$ imparts a momentum kick due to the photon recoil after the emission of a photon with polarization $q$ in direction $\hat{\bm{n}}$.  In general, the recoil operator is averaged over the angular emission profile for each polarization.  If we assume that the photon is emitted along either direction of the $x$-axis, then the recoil operators are $R_{\pm,q} = e^{\pm ikx}$, and we have a total of six collapse operators for the two emissions directions and three polarizations
\begin{equation}
    {L}_{\pm,q} = \sqrt{\Gamma} \, e^{\pm i k \hat{x}} \hat{D}_q.
\end{equation}

In 1D, the photon recoil operator can be understood as a momentum translation operator that induces a momentum kick due to the recoil from an emitted photon, with  $e^{\pm i k \hat{x}} |\psi(p) \rangle =  |\psi(p \pm \hbar k) \rangle$.  

We can express the position operator in terms of the harmonic oscillator creation and annihilation operators $\hat{x} = x_0 ( \hat{a} + \hat{a}^\dag )$, where $x_0 = \sqrt{\hbar/ 2 m \nu}$ is the width of the ground state wavefunction. The recoil operator becomes   $ e^{i k \hat{x}} = e^{i \eta ( \hat{a} + \hat{a}^\dag)}$, where the Lamb-Dicke (LD) parameter $\eta = k x_0$ is the ratio of the ground state wavefunction width to the wavelength  {of the cooling light}. 

The square of the LD parameter also happens to be equal to the ratio of the photon recoil energy $E_\text{recoil} = \hbar^2 k^2/ (2m) $ to the harmonic energy $\hbar \nu$, or $\eta^2 = E_\text{recoil}/(\hbar \nu)$. 
When $\eta \ll 1 $, then the recoil from photon emission or absorption is unlikely to change $n$,  {in addition to the ground state function being} \del{and the ground state wavefunction is} smaller than the cooling laser's wavelength. One common misconception is that in the LD regime, heating is suppressed. Although the probability of $n$ changing decreases, the overall heating rate is still given by the scattering rate times the recoil energy.  In this regime, we can also approximate the recoil operator  {to first order in the LD parameter} as 
\begin{equation} \label{eq:L.D. expansion}
 e^{ i k \hat{x}} \approx 1 + i \eta (\hat{a} + \hat{a}^\dag) .  
\end{equation}
The result of a recoil is then  {to transfer population to neighboring motional states}
\begin{equation}
    e^{ i k \hat{x}} |n \rangle \approx  |n \rangle + \eta \sqrt{n} |n-1 \rangle + \eta \sqrt{n+1}|n+1 \rangle
\end{equation}
The zeroth order term is from $n \rightarrow n$, and the first order LD term is from $n \rightarrow n \pm 1 $. The LD regime is defined as when \( \eta^2 (2n+1) \ll 1 \).  

\del{More generally,}  {Outside the LD regime,} the matrix element  { of the recoil operator between different motional states }can be expressed in terms of the generalized Laguerre polynomial as~\cite{wineland1979laser} 
\begin{gather}
    \langle n' | e^{i k \hat{x}} | n \rangle = \\
    e^{-\eta^2/2} \left(\frac{n_<!}{(n_< + \Delta n)!}\right)^{1/2}      (i \eta )^{\Delta n}   \, L_{n_<}^{\Delta n}(\eta^2)  \nonumber .
\end{gather}
Here, $n_<$ is the least of $n$ and $n'$, and $\Delta n = |n'-n|.$  
This functional form becomes important when cooling outside the LD regime. The $n \rightarrow n-1$ transition goes to zero for higher $n$, creating dark motional states where the population gets stuck.  Cooling can still occur by addressing higher order sidebands, as shown in Ref.~\cite{ Morigi1999Laser, Roghani2008Trapped, yu2018motional}. 

The master equation here is simulated throughout this paper for different cooling schemes using the steady-state solver in QuTip~\cite{qutip}.  We include up to 15 harmonic states, which for the larger spin systems in this paper like $F=1$ to $F'=2$ takes approximately one minute per simulation. 

\subsection{Effective master equation within ground state subspace} \label{subsec:effective}
\del{Most optical cooling processes take place when there is low saturation}  {Optical cooling typically takes place in the low saturation regime}, meaning that only a small amount of the steady-state population is in the excited state. The excited state can be\del{reasonably}  {accurately} approximated from the ground state, and can be adiabatically eliminated  {to create effective ground state dynamics}. In this section, we will derive the effective ground state master equation.

In situations where the excited state quickly reaches an equilibrium, the excited state can be solved in terms of the ground states and then eliminated from the time evolution equations. The elimination results in an effective Hamiltonian and collapse operator that gives the dynamics within the ground state.  These effective operators can be derived by performing these operations for a given Hamiltonian~\cite{deutsch2010quantum}, but Ref.~\cite{reiter2012effective} provides a compact general formalism that we use here.  

 {Most generally, for light that is either close to resonance or far-from resonance, the effective ground state Hamiltonian and collapse operators can be expressed as a sum over the ground and excited spin states as~\cite{reiter2012effective}}
\del{This first form is valid even when the cooling light is close to resonance such as in sideband cooling.  Hamiltonian and collapse operators within the ground state subspace are}  
\begin{gather}
H_\text{eff} = \nu a^\dag a \nonumber \\ 
-\frac{1}{2} \left[ \sum_{m_g,m_e} \frac{P_g H |m_e \rangle \langle m_e| H|m_g \rangle \langle m_g |   }{( \Delta_{m_e, m_g} - i \Gamma/2 )}  
+ h.c. \right]     \label{eq:Hclose1}
\end{gather}
\begin{gather}   \label{eq:Lclose1}
L_\text{eff}^i =  L_i \sum_{m_g, m_e} \frac{ |m_e \rangle \langle m_e|  H  | m_g \rangle \langle m_g | }{\Delta_{m_e, m_g} - i \Gamma/2} .
\end{gather}
The laser detuning  {between two specific spin states} is $\Delta_{m_g,m_e} = \omega_L - (E_{m_e} - E_{m_g} ) $.   {The first and last operators in both of these expressions operate only on ground state spins.}

 {Substituting the Hamiltonian for a spin interacting with a single laser frequency from Eq.~\ref{eq:H} gives an effective Hamiltonian}
\del{For the case of an atom with spin interacting with a single frequency laser field, as in the Hamiltonian in Eq.~\ref{eq:H}, the effective ground state Hamiltonian becomes}
\begingroup
\small
\begin{gather} \label{eq:Hclose}
H_\text{eff} = \nu \hat{a}^\dag \hat{a} - \\
\frac{\Omega^2}{8} \! \left[ 
  \sum_{m_g,m_e} \!\! \frac{ (\bm{\epsilon}
(\hat{x}) \! \cdot \! \bm{D}) |m_e \rangle \langle m_e | (\bm{\epsilon}^*(\hat{x}) \! \cdot \! \bm{D}^\dag) |m_g \rangle \langle m_g |}{\Delta_{m_g,m_e} - i \Gamma/2 }   
  + h.c. \right]   \nonumber
\end{gather}
\endgroup
and the collapse operator for the emission of a photon with polarization $q$ becomes
\begingroup
\small
\begin{gather} \label{eq:Lclose}
L_\text{eff}^{q, \hat{\bm{n}}} = \! - \sqrt{\Gamma} \, \frac{\Omega}{2} \!\! \sum_{m_g, m_e} \!\! \frac{ (\mathcal{R}_{\bm{n}} \hat{\bm{e}}_q^* \cdot \bm{D})|m_e \rangle \langle m_e| (\bm{\epsilon}(\hat{x}) \cdot \bm{D}^\dag)| m_g \rangle \langle m_g |}{ \Delta_{m_g,m_e} - i \Gamma/2} .
\end{gather}
\endgroup  
The effective Hamiltonian includes light shifts, two-photon Raman coupling, and decay due to spontaneous emission. The collapse operators repopulate the ground state  {after spontaneous emission}.  {These expressions are valid close to resonance and will be used for resolved and unresolved sideband cooling.}

Two photon cooling schemes typically operate in the large detuning limit. In the large detuning limit, \del{the state dependence in the denominator detuning vanishes}  {the state-dependent detuning can be replaced by a constant detuning} $\Delta_{m_g, m_e} \approx \Delta$. The sum over the excited states in the numerator  {in both operators} becomes the identity. The effective ground state operators from Eqs.~\ref{eq:Hclose1} and \ref{eq:Lclose1} simplify to
\begin{gather} \label{eq:Hproj}
H_\text{eff} = \nu a^\dag a -\frac{1}{2 \Delta} \left( P_g H P_e  H P_g    %- i \Gamma/2, + h.c.
 \right)
\end{gather}
\begin{gather} \label{eq:Lproj}
L_\text{eff}^q =  L_q \frac{P_e H  P_g }{\Delta - i \Gamma/2}. 
\end{gather}
%A similar expression also follows from 2nd order perturbation theory.
 {The single frequency spin operators from Eqs.~\ref{eq:Hclose} and \ref{eq:Lclose}} become
\begin{gather} \label{eq:Hdetuned}
H_\text{eff} = \nu \hat{a}^\dag \hat{a} -
\frac{\Omega^2}{4 \Delta} \left[
 (\bm{\epsilon}^*(\hat{x}) \cdot \bm{D}) (\bm{\epsilon}(\hat{x}) \cdot \bm{D}^\dag)\right]   % - i \Gamma/2
\end{gather}
\begin{gather} \label{eq:Ldetuned}
L_\text{eff}^{q,\hat{\bm{n}}} = - \sqrt{\Gamma} \, \frac{\Omega}{2} \left[  \frac{ ( \mathcal{R}_{\hat{\bm{n}}} \hat{\bm{e}}_q^* \cdot \bm{D}) (\bm{\epsilon}(\hat{x}) \cdot \bm{D}^\dag) }{ \Delta - i \Gamma/2}   \right].
\end{gather}

The \del{interaction part of the}ground state Hamiltonian describes two-photon processes such as light shifts and Raman transitions.  The right term in the numerator 
$\bm{\epsilon}(\hat{x}) \cdot \hat{\bm{D}}^\dag$ excites the atom due to absorption of a photon from the cooling light. The left term $\bm{\epsilon}^*(\hat{x}) \cdot \hat{\bm{D}}$ returns the atom to the ground state with the emission of a photon into the cooling light. 
 
The ground state collapse operator describes the redistribution of population within the ground state due to spontaneous emission. It has a similar form as the interaction Hamiltonian. The right term $\bm{\epsilon}(\hat{x}) \cdot \hat{\bm{D}}^\dag$ excites the atom due to the absorption of a photon from the cooling light.  The left term $ \mathcal{R}_{\bm{n}} \bm{e}_q^* \cdot \hat{\bm{D}}$ returns the atom to the ground state by emitting a photon into free-space and gives the atom a momentum kick through the recoil operator $\mathcal{R}_{\bm{n}}$. 

 {  The operators operate within the ground state spin manifold.  Therefore, we can also express these operators in terms of the spin operators $\hat{F}_{i}$. While the derivation is long, the idea is quite simple.  After adiabatic elimination, both operators are in the form }
\del{After adiabatic elimination, the form of both the effective Hamiltonian and collapse operators are in terms of the scalar products of the dipole operators and field polarization, }
\begin{equation} \label{eq:convert_spin_to_operators}
   (\bm{A}^* \cdot \hat{\bm{D}} ) ( \bm{B} \cdot \hat{\bm{D}}^\dag  )  = (\bm{A}^* \otimes \bm{B}) \cdot (\hat{\bm{D}} \otimes \hat{\bm{D}}^\dag)  . 
\end{equation}
 {Here $\bm{A}$ and $\bm{B}$ are vectors related to the electric field of the absorbed or emitted photons. 
 On the right side we re-express it  as a scalar product of two rank 2 tensors.  The operator $\hat{\bm{D}} \otimes \hat{\bm{D}}^\dag$ is a rank 2 tensor, formed by the two tensor product of two rank 1 vectors.  This rank 2 operator can be decomposed into an irreducible representation which consists of a rank 0 scalar, rank 1 vector, and rank 2 symmetric and traceless tensor. Finally, the Wigner-Eckert theorem can be used on each of these components to convert them to spin matrix elements. 
 }
 \del{
These operators act within the ground state spin manifold just like the hyperfine spin operators $\hat{F}_i$, $F_{\pm}$. If you observe, the right hand side is actually a scalar product of two rank 2 matrices.  Using the Wigner-Eckart theorem and properties of spherical vectors,  we can express these in terms of spin operators.}  The spin Hamiltonian becomes~\cite{deutsch2010quantum}
\begin{gather} \label{eq:Hdetuned2}
H_\text{eff} = \nu \hat{a}^\dag \hat{a} -
\frac{\Omega^2}{4 \Delta} \left[  C^{(0)} | \bm{\epsilon}(\hat{x})|^2   \right.  \\
\left. + C^{(1)} i [\bm{\epsilon}^*(\hat{x}) \times \bm{\epsilon}(\hat{x})] \cdot \hat{\bm{F}}  \right.  \nonumber \\
\left.+ C^{(2)} \left(   \frac{1}{2}[  (\bm{\epsilon}(\hat{x}) \!\cdot\! \hat{\bm{F}})^\dag (\bm{\epsilon}(\hat{x}) \!\cdot\! \hat{\bm{F}}) + c.c. ]  - \frac{1}{3} \hat{\bm{F}}^2 |\bm{\epsilon}(\hat{x})|^2 \right) \right] \nonumber . 
\end{gather}  
 {The coefficients $C^{(i)}$ are given in an appendix.}
The collapse operators which give the redistribution of population in the ground state due to spontaneous emission are
\begin{gather}  \label{eq:Ldetuned2}
L_\text{eff} = -\sqrt{\Gamma} \frac{\Omega}{2 (\Delta - i \Gamma/2)} \left[ 
C^{(0)}  (\mathcal{R}_{\bm{n}} \hat{\bm{e}}_q^*) \cdot \bm{\epsilon}(\hat{x})  \right. \\
\left. 
+ C^{(1)} i ([(\mathcal{R}_{\bm{n}} \hat{\bm{e}}_q^*) \times \bm{\epsilon}(\hat{x})] \cdot \hat{\bm{F}})  \right. \nonumber \\
+ C^{(2)} \left [ \frac{1}{2}[  (\mathcal{R}_{\bm{n}} \hat{\bm{e}}_q^*) \! \cdot \! \hat{\bm{F}})^\dag (\bm{\epsilon}(\hat{x}) \!\cdot\! \hat{\bm{F}}) + h.c. ]  \right. \nonumber \\
\left. - \frac{1}{3} \hat{\bm{F}}^2 |(\mathcal{R}_{\bm{n}} \hat{\bm{e}}_q^*) \cdot \bm{\epsilon}(\hat{x})| \right]  \nonumber  .
\end{gather}
 The effective spin operators are valid for any ground and excited spin states. Interestingly, the excited state spin only affects the $C^{(i)}$ coefficients. 

%\subsection{Population rate equation} \label{subsec:population}
The ground state collapse operators in Eqs.~\ref{eq:Ldetuned} (far-detuned limit) and \ref{eq:Lclose} (near-resonant limit) describe the redistribution of the population within the ground state. With the effective ground state Hamiltonian, a master equation incorporating these operators describes the ground state dynamics. The Hamiltonian accounts for energy shifts and coherent transfer between states, while the collapse operators account for incoherent redistribution due to spontaneous emission.

Next, we will derive an approximate population rate equation  {for the ground state spin levels} by working in the basis where \(H_\text{eff}\) is diagonal.  {While \(H_\text{eff}\) captures the coherent dynamics within the ground state, the collapse operators capture population redistribution from spontaneous emission.}    We define an interaction picture by applying a unitary transformation \(\tilde{A} = U A U^\dag\) with \(U = e^{-i H_{\text{eff}} t }\). In the interaction picture, the time evolution of the Hamiltonian is included within the density matrix and operators \(U \rho U^\dag\). In the interaction picture, the master equation is:
\begin{equation}
\dot{\tilde{\rho}} =  \sum_q \left[ \tilde{L}_{\text{eff},q} \, \tilde{\rho} \, \tilde{L}_{\text{eff},q}^\dag - \frac{1}{2} \{ \tilde{L}_{\text{eff},q}^{\dag} \tilde{L}_{\text{eff},q} , \, \tilde{\rho} \} \right].
\end{equation}

Finally, we derive rate equations for the populations by multiplying by \(\langle n |\) on the left and \(| n \rangle\) on the right and defining the populations \(P(n) = \tilde{\rho}_{nn} = \rho_{nn}\). The resulting equation contains coherence terms \(\langle n | \rho | m \rangle\) between ground states. If we work in the basis where the ground state Hamiltonian is diagonal, these coherences are less important  and can often be ignored. By setting all these coherences to zero, we obtain a population rate equation:
\begin{equation} \label{eq:rate}
\frac{dP(n)}{dt} = - P(n) \sum_{m} \gamma_{n \rightarrow m} + \sum_{m} \gamma_{m \rightarrow n} P(m),
\end{equation}
where the transfer rate between ground states \(n\) and \(m\) is:
\begin{equation} \label{eq:rateL}
\gamma_{n \rightarrow m} =  \sum_q |\langle m | L_{\text{eff},q} | n \rangle|^2.
\end{equation}
Note that in this last equation, \(L_\text{eff}\) can be taken out of the interaction picture because as long as \(n\) and \(m\) are eigenstates of \(H_\text{eff}\), the time-dependence cancels upon taking the magnitude. However, it is important to remember that the eigenstates must be those of \(H_\text{eff}\).
The steady-state conditions can then be found by solving the matrix equation for \(\frac{dP(n)}{dt} = 0\). In this paper, we derive analytic solutions for different cooling schemes by first picking an appropriate basis where the coherences between eigenstates are least important and then \del{by}solving this rate equation.

To compare different cooling schemes we need to set some parameter as a metric. The energy or temperature of a single atom confined in a 1D harmonic potential is given by $E = \hbar \nu(\langle n\rangle + 1/2)$, where $\nu$ is the trapping frequency, and $\langle n\rangle$ is the average motional quantum number.  However, it is important to note that the energy of a single atom in an optical tweezer can be manipulated by adiabatically varying the optical trap power~\cite{tuchendler2008energy}, where both the trapping frequency and the energy scale with the square root of the optical trap power. As a result, the motional quantum number, $\langle n\rangle$, becomes a more relevant quantity to consider, as it remains conserved during adiabatic changes in the trap depth. Therefore, we choose to use $\langle n\rangle$ as the cooling metric for comparing all the techniques in the following discussions.

%Experiments often quote temperature, as that is more directly obtained through release and recapture measurements where the trap is abruptly turned off and on for a short time and the probability of survival is measured, corresponding to the distribution of motional quantum numbers, $n$, within the trap. 

%Release-recapture is an experimental technique used to determine the temperature of trapped atoms by assuming a Boltzmann distribution for a Gaussian harmonic potential. In this method, the atom is released from the trap for a finite duration, allowing it to expand freely. By measuring the kinetic energy of the atom upon its release and comparing it to the theoretical predictions based on the assumed energy distribution, researchers can infer the temperature of the atom. This temperature corresponds to the distribution of motional quantum numbers, $n$, within the trap. While the release-recapture technique provides a practical means to estimate the trapped atom's temperature, it relies on the assumption of a specific energy distribution and the ability to accurately measure the atom's kinetic energy after release. Despite these limitations, this method remains a widely used tool for characterizing the energy distribution of trapped atoms in experimental settings.

%%%%%%%%%%%%%%%%%%%%%%%%%%%%%%%%%%%%%%%%%%%%%%%%%%%%%%%%%%%%%%%%%%%%%
\section{Sideband cooling}   \label{sec:sideband}
\begin{figure*}[t]
  \centering
  \includegraphics[width=0.99\linewidth]{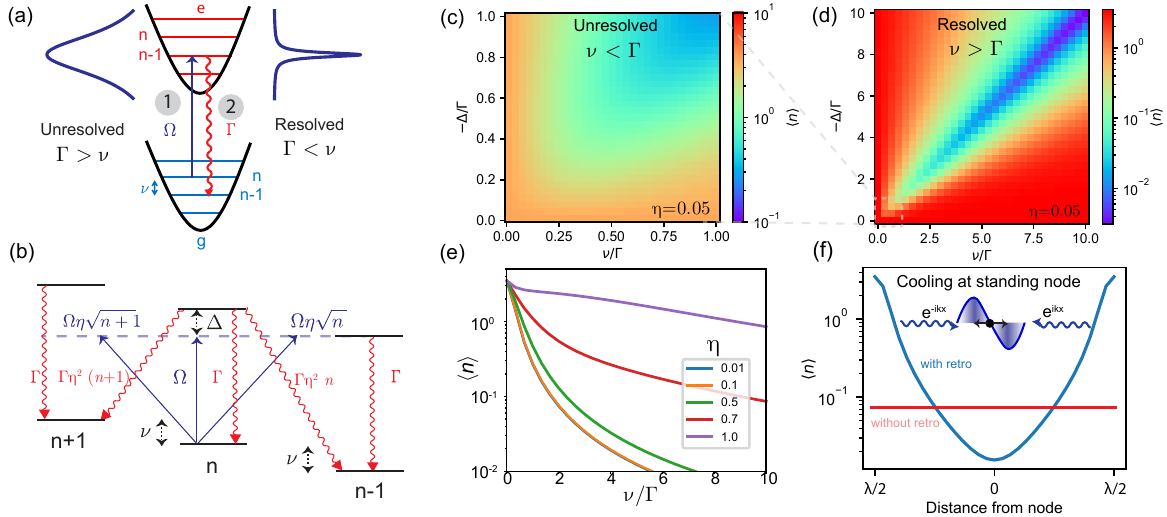}
  \caption{ \textbf{(a)} Resolved ($\Gamma < \nu$) and unresolved ($\Gamma > \nu$) sideband cooling in the Lamb-Dicke regime.  {A red-detuned laser drives $n$ to $n-1$, which mostly decays down to the $n-1$ ground state.}
  \textbf{(b)} The cooling laser drives the three blue transitions: $n \rightarrow n$ and  $n \rightarrow n \pm 1$. Spontaneous emission (red) most likely does not change $n$ in the LD regime.   \textbf{(c-d)} Simulation of $\langle n \rangle $ in the unresolved and resolved regime for a LD parameter $\eta = 0.05$. \textbf{(e)} The final population versus trapping frequency for varying $\eta$.  {The lines converge below $\eta =0.1$.} \textbf{(f)}  Cooling in a standing wave can lower the energy by positioning the atom at the node of the field, thereby decreasing scattering but still allowing sideband cooling due to the presence of a gradient field.  }
  \label{fig:Doppler}
\end{figure*}
% c,d =  eta = 0.05, Nh = 8, Omega1=sqrt(.05), delta1 = -scan[1]. 
% e = Nh = 8, J=0 to J=1, lin, Omega=sqrt(0.05), delta = -wh.  
%f:   wh = 2, .05, 8, lin_retro, Omega1=ssqrt(0.05), delta=-wh, phi = scan*pi.
%
%Doppler cooling in free space relies on the relativistic Doppler shift of an atom being brought closer to resonance when moving toward a red-detuned laser. The final Doppler temperature $T_\text{Doppler} = \hbar \Gamma/2k_B $ results from a balance of the decrease in momentum of $\hbar k$ from the absorption of the cooling laser and the recoil heating $E_R = \hbar^2 k^2/ 2 m $ that results from spontaneous emission in all directions.  Here $\Gamma$ is the excited state linewidth, $k_B$ is the Boltzmann constant, $k$ is the wave-vector of the cooling light, and $m$ is the mass of the atom.

 {In this section, we start looking at specific cooling schemes, starting with sideband cooling, a single-photon cooling schemes that works even for a two-level system.  
In free-space Doppler cooling, an atom moves towards a red-detuned light source, and the Doppler effect brings them closer to resonance, causing them to absorb photons and receive momentum kicks that oppose the motion. }

 {
For an atom confined in a harmonic trap within or close to the Lamb-Dicke (LD) regime, a different picture arises, as illustrated in Fig.~\ref{fig:Doppler}(a).} In this scenario, a red-detuned cooling laser drives transitions from the motional state $n$ to $n-1$, while spontaneous emission predominantly does not change $n$, returning the atom to the ground $n-1$ state. This process leads to net cooling because, in the LD regime, the recoil energy from photon emission is smaller than the energy spacing between trap levels ($\hbar \nu$).

Doppler cooling of trapped atoms can be classified into two regimes based on the relationship between the trapping frequency $\nu$ and the atomic transition linewidth $\Gamma$:
\begin{itemize}
    \item Resolved sideband cooling: when $\nu > \Gamma$
    \item Unresolved sideband cooling: when $\nu < \Gamma$
\end{itemize}

These regimes exhibit distinct cooling dynamics and efficiencies.
Sideband cooling was extensively studied theoretically in the 1980's by
Refs.~\cite{wineland1979laser, javanainen1981laser, stenholm1986thesemiclassical, leibfried2003quantum}. It has been demonstrated experimentally in ions~\cite{peik1999sideband}, neutral atoms~\cite{perrin1998sideband, han20003d}, and nanomechanical oscillators~\cite{schliesser2008resolved}.  % check references
Here we present these results in our formalism.  {Unlike for ion traps, optical traps typically have different potentials for the ground and excited states. We extend this formalism to treat different trapping frequencies and derive an analytic expression for the steady-state temperature as a function of the mismatch.} \del{We then look at the case of different trapping frequencies and derive an analytic expression for the steady-state temperature.}

\subsection{Sideband cooling for the same ground and excited trapping frequencies}
Doppler cooling for a bound atom in the LD regime only requires a single traveling wave $\bm{\epsilon}(\hat{x}) = e^{i k \hat{x}} $.  For a two-level system, the dipole operators are $\hat{D} = \hat{\sigma}  = |g \rangle \langle e|$. The Hamiltonian is 
\begin{equation} \label{eq:sbH}
    H = \nu \hat{a}^\dag \hat{a} +  \Delta |e \rangle \langle e| - \frac{\Omega}{2}  \left( e^{i k \hat{x}}  \hat{D}^\dag + h.c.  \right), 
\end{equation}
and the collapse operators for emission into the positive and negative $x$-directions  are
\begin{equation}  \label{eq:sbL}
    {L}_{\scriptscriptstyle  \pm} = \sqrt{\Gamma} \, e^{\pm i k x} \hat{D} .
\end{equation}

The resonant and spontaneous emission processes are shown in Fig.~\ref{fig:Doppler}(b).  {In LD regime, the population in the ground $n$ state is driven to excited $n$ and $n \pm 1$.  The matrix element to $n-1$ is reduced by $\eta \sqrt{n}$ relative to same $n$.  However, as shown in (b), the transition can be preferred if made resonant by detuning the laser by $\Delta = \nu$.   The excited then decays back to the same $n$ with rate $\Gamma$, and changing $n$ with reduced rate $\Gamma \eta^2 n$.}   To estimate the final population distribution, we can use the effective ground state collapse operator in Eq.~\ref{eq:Lclose} and the population transfer rates from Eq.~\ref{eq:rateL}.  From these equations, the transfer rate between ground states $n_1$ and $n_2$ in the ground state through the excited states $n_e$ is
\begin{equation} \label{eq:dopplerrate}
    \gamma_{n_1 \rightarrow n_2} =  \Gamma \,\, \Omega^2  \sum_{\pm}  \left|  \sum_{n_e} \frac{ \langle n_2 | e^{\pm i k \hat{x}} | n_e  \rangle  \langle n_e|  e^{i k \hat{x}} | n_1 \rangle}{ \Delta + \nu (n_1 - n_e  ) - i \Gamma/2 }  \right|^2  .
\end{equation} 
 
The sum inside the magnitude is over all excited states $n_e$.  {Pathways from $n_1$ to $n_2$ ground states can in principle interfere through these excited states. }The sum on the outside is over the emission direction. 
 {The random emission into different directions results in a random phase that ends up averaging out the coherence terms, suppressing the interference.} \del{Emission into each direction results in a different phase, and the average over all directions averages out the coherences between the different pathways.}As a result, we can take the sum over the excited states outside of the sum and omit the average over directions, 
\begin{equation} \label{eq:dopplerrate2}
    \gamma_{n_1 \rightarrow n_2} =  \Gamma \,\, \Omega^2 \sum_{n_e}  \left|   \frac{ \langle n_2 | e^{ i k \hat{x}} | n_e  \rangle  \langle n_e|  e^{i k \hat{x}} | n_1 \rangle}{ \Delta + \nu (n_1 - n_e  ) - i \Gamma/2 }  \right|^2  .
\end{equation} 
This rate describes two processes. The first term in the numerator $|\langle n_2 | e^{ik\hat{x}} | n_e\rangle|^2$ describes the process of recoil due to spontaneous emission. The rest of the expression gives the scattering rate of photon absorption.   Together, these two terms give the total rate of photon absorption and emission into motional states and the resulting recoil.    

In the LD regime, the allowed motional state transition rates are
\begin{gather} \label{eq:ApAm}
    \gamma_{n \rightarrow n+1} =  \eta^2 (n+1) \, A_{\scriptscriptstyle +} \\
    \gamma_{n \rightarrow n-1} =  \eta^2 n \, A_{\scriptscriptstyle -}   \nonumber
\end{gather}
  where $A_{\scriptscriptstyle +}$ and $A_{\scriptscriptstyle -}$ are 
\begin{gather}
    A_{\scriptscriptstyle +} =   R(\Delta - \nu)  + \alpha R(\Delta )  \\
    A_{\scriptscriptstyle -} =   R(\Delta + \nu)   +  \alpha R(\Delta ).  \nonumber
\end{gather}
The function $R(\Delta)$ is the low-saturation two-level system scattering rate
\begin{equation}
    R( \Delta )  = \Gamma \, \frac{\Omega^2/4}{ \Delta^2+(\Gamma/2)^2 } .  % see 5.277 on pg 197 of Steck.
\end{equation}
The constant $\alpha$ is related to the averaging over emission directions.  Assuming emission into 1D gives $\alpha=1$, but averaging over emission into all directions leads to $\alpha = 1/3$~\cite{javanainen1981laser}.

In the LD regime, transfer only occurs between neighboring states, in which case the rate equation from Eq.~\ref{eq:rate} is 
\begin{gather} \label{eq:Pnd2}
    \dot{P}(n) = \eta^2 \, A_{\scriptscriptstyle -} \, (n+1)\, P(n+1)  + \eta^2 A_{\scriptscriptstyle +} \, n  \, P(n-1)  \nonumber \\ 
    -   \eta^2 \,A_{\scriptscriptstyle +}  \,  (n+1)\, P(n) -   \eta^2  A_{\scriptscriptstyle -} \, n\, P(n) . 
\end{gather}
 The cooling laser (Fig.~\ref{fig:Doppler}(b), blue) drives the $n+1$ and $n-1$ sidebands with effective Rabi frequencies $\Omega \eta \sqrt{n+1}$ and $\Omega \eta \sqrt{n}$. Decay (Fig.~\ref{fig:Doppler}(b), red) of excited state to sideband occurs with probability $\Gamma \eta^2 n$. 

 {Next we calculate the steady-state temperature.}
Because the population only transfers between neighboring $n$ in the LD regime, we can calculate the steady-state population by assuming no flow of population between $n$ and $n+1$, which gives
\[P(n+1) \, \gamma_{n+1 \rightarrow n} = P(n) \,\gamma_{n \rightarrow n+1}.\] 
Solving this equation  {and substituting Eq.~\ref{eq:ApAm}} leads to the steady-state condition
\begin{equation}
    \frac{P(n+1)}{P(n)} = \frac{A_{\scriptscriptstyle +}}{A_{\scriptscriptstyle -}}.
\end{equation}
We can set this equal to the Boltzmann factor to get the steady-state temperature:
\begin{equation}
    \frac{P(n+1)}{P(n)} = \exp{\left(-\frac{\hbar \nu }{k_B T} \right) }.
\end{equation}
This represents a geometric distribution.  Defining this ratio as $s$, the normalized solution is $P(n) = (1-s) s^n$, and the expectation and variance of $n$ is $\langle n\rangle = \frac{s}{1-s}$ and $ \langle (\Delta n)^2 \rangle  = \frac{s}{(1-s)^2}$.

 {We can also derive an expression for the time evolution of the average population.}
Multiplying Eq.~\ref{eq:Pnd2}  by $n$ and summing over $n$ gives the time rate equation for $\langle n \rangle$:
\begin{gather}
    \frac{d\langle n \rangle}{dt} 
    =   \eta^2 \left[ A_{\scriptscriptstyle  +} (\langle n \rangle + 1) - A_{\scriptscriptstyle -} \langle n \rangle  \right]
\end{gather}
The cooling rate $ \eta^2 (A_{\scriptscriptstyle -} - A_{\scriptscriptstyle +}) $ is balanced by the constant heating from spontaneous emission $\eta^2 A_{\scriptscriptstyle +}$.
The steady-state solution is 
\begin{gather} \label{eq:nssA}
    \langle n \rangle  % = \left( e^{- \hbar \nu / k_B T} - 1\right)^{-1}  \\
    =  \frac{A_{\scriptscriptstyle +}}{ A_{\scriptscriptstyle -} - A_{\scriptscriptstyle +}}  = \frac{R(\Delta- \nu) + \alpha R(\Delta)}{ R(\Delta+\nu) - R(\Delta - \nu) }. 
\end{gather}
In the resolved-sideband regime ($\Gamma \ll \nu$), the lowest temperature occurs at the cooling sideband $\Delta = - \nu$, at which the steady-state is 
\cite{wineland1979laser, stenholm1986thesemiclassical} 
\begin{equation} \label{eq:resolvedtheory}
   \langle n \rangle =    \frac{1}{4} \left(\frac{\Gamma}{\nu}\right)^2 \left(\alpha + \frac{1}{4}\right) .
\end{equation}
In the unresolved-sideband regime ($\Gamma \gg \nu$), the steady state is 
\begin{equation}
    \langle n \rangle = \frac{(1 + \alpha)}{8} \left(  \frac{\Gamma}{2 \Delta}  +  \frac{2 \Delta}{\Gamma} \right)  - \frac{1}{2}.
\end{equation}

 {In Fig.~\ref{fig:Doppler}, we simulate sideband cooling with the  master equation from Eqs.~\ref{eq:sbH} and \ref{eq:sbL}.   We use the steady-state solver in QuTip using up to 15 harmonic states in both the ground and excite state and assume 1D spontaneous emission.  }
\del{In Fig.~\ref{fig:Doppler} we simulate the full master equation for the ground and excited states. }
The simulations in Fig.~\ref{fig:Doppler}(c-d) are in the unresolved and resolved sideband regimes for $\eta = 0.05$ and in the low saturation limit $\Omega = 0.05 \, \Gamma$. In the resolved regime in (d), best cooling occurs at $\Delta = -\nu$ and agrees closely with Eq.~\ref{eq:resolvedtheory}.  The unresolved regime in (c) is a zoomed in plot of the lower left corner of (d).  Theoretically in the unresolved regime, the lowest temperature occurs for $\Delta = -\Gamma/2$, although the exact solution shows that the optimum is broad and not much better than the resolved criteria $\Delta = -\Gamma$.

In Fig.~\ref{fig:Doppler}(e), we plot steady-state population versus the trapping frequency for various LD parameters $\eta$.   {  The population converges for $\eta < 0.1$, with the last two curves $\eta = 0.1$ and $\eta = 0.01$ nearly identical.  These curves agree well with Eq.~\ref{eq:nssA} for $\alpha = 1$.  From this, we can see that getting 99\% ground state ( or $\langle n \rangle  = 10^{-2} $) requires $\nu / \Gamma  > 5$ for $\eta < 0.1$.  Typical trapping frequencies in optical tweezers are $\sim 100 $ kHz. Because alkali and alkaline-earth linewidths are 5 MHz for the nS to nP transition, sideband cooling cannot prepare ground state atoms.
}

\subsection{Dark sideband cooling}
 {One lesson from the sideband cooling formalism is that the sideband is driven by the gradient of the field, not the field itself.}   This can be seen by expanding the atom-field interaction as a series to first order in terms of the LD parameter and field gradient 
\begin{equation}
  H_I =   -\frac{\Omega}{2} \left[ \left( \bm{\epsilon}(0) + \hat{x} \frac{d\bm{\epsilon}(0)}{d \hat{x}} \right) \cdot \bm{D}^\dag  +h.c. \right].
\end{equation}
 {This gradient can be a complex phase gradient, for example due to a traveling wave with $e^{ikx} \approx 1 + i k \hat{x} $, or an amplitude gradient, for example from a standing wave $\sin(kx) \approx kx$.  But that means that a field at the atom equilibrium position is not necessary, and even detrimental as it drives the $n \rightarrow n$ transition with a resulting recoil.   }
The cooling performance of sideband cooling can be improved by adding a second counter-propagating beam with the same polarization to form a standing wave and placing the atom at the node of the field~\cite{cirac1993laser}.  Although the atom now sees zero intensity at the equilibrium position, the sidebands are still driven.
\del{The gradient can be either a phase or amplitude gradient, both of which drive the sideband. Intensity at the position of the atom does not lead to cooling and, in fact, only leads to extra heating.}  If we place the atom in a standing wave with linear polarization, the field is 
\(    \bm{\epsilon}(\hat{x}) = \sin(k \hat{x} )  \). 
Fig.~\ref{fig:Doppler}(f) simulates the steady-state population versus atom position with $\nu$ = $-\Delta$.  {The population reduces by a factor of 5 at the standing wave node.} Theoretically, this gets rid of the process in Fig.~\ref{fig:Doppler}(b) of driving $n$ to $n$, followed by spontaneous emission heating to the sideband.  {The steady-state solution in Eq.~\ref{eq:nssA} ends up being the same except with $\alpha = 0$.} 

Cooling can be further enhanced by placing the atom in a cavity that can enhance the red sideband~\cite{vuletic2001three, David2009Cavity}.  {A similar idea could also in principle be obtained by shaping the light mode profile, for example placing the atom at the dark center of a donut beam where a radial gradient still exists, although this has not been theoretically or experimentally investigated.}

\subsection{Sideband cooling for excited state with different trapping frequency} \label{subsec:squeezing}
\begin{figure*}[tb]
  \centering
  \includegraphics[width=0.9\linewidth]{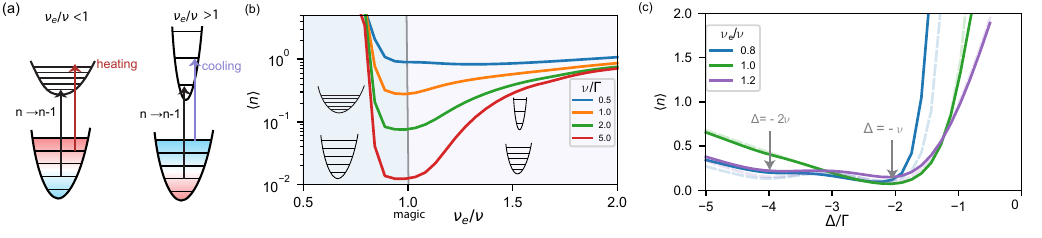}
  \caption{ Sideband cooling with different excited $\nu_e$ and ground state $\nu$ trapping frequencies. \textbf{(a)} Two regimes, where the excited state is less trapped (left) or more trapped (right) result in extra heating.  This heating comes from coupling many different motional states in the quantum picture, or a fluctuating dipole force in the classical picture.  For different trapping frequencies, the $n$ to $n-1$ transition can be resonant for only one specific $n$.  For $\nu_e < \nu$, higher $n$ gradually shift more blue towards heating.  For $\nu_e > \nu$, higher $n$ gradually shifts more red, leading to cooling of higher order sidebands.   \textbf{(b)} Population for a different excited state trapping frequency $\nu_e$, plotted for different $\nu/\Gamma$ and $\eta =0.1$ and $\Delta = - \nu$. The magic condition is $\nu_e/\nu=1$.  \textbf{(c)} Population versus detuning for different excited to ground trapping frequency ratios $\nu_e/\nu$. A minimum occurs at $\Delta = -\nu$, but a second minimum occurs at $\Delta = - 2 \nu$ when $\nu_e \ne \nu$. Plot is for $\eta=0.1$, $\nu/\Gamma = 2$. The analytical model in Eq.~\ref{eq:squeezinganalytical} is plotted in a light dashed line. Besides heating, different trapping frequencies also allows the coupling of higher order sidebands, providing new opportunities for cooling.}  % both are \Omega/\Gamma=\sqrt(.05).  $N_h=12$ for a, 9 for b. 
  \label{fig:Squeezing}
\end{figure*}

The Doppler cooling section assumed that the electronic ground and excited states have the same trapping frequency.  But typically, the ground and excited states of neutral atoms have different polarizabilities except at ``magic wavelengths."   {This is due to the excited state being coupled to several higher energy states.} The different potentials lead to heating due to the abruptly changing dipole force during absorption or emission, shown in the schematics in Fig.~\ref{fig:Squeezing}.  In the extreme case, the excited state is anti-trapped resulting in loss of the atom if the excited lifetime comparable to the oscillation period $1/\nu$.  

Because magic wavelengths are not often convenient for optical trapping, neutral atoms more often use two-photon cooling schemes, such as polarization gradient cooling and Raman sideband cooling. 
However, alkaline-earth atoms contain narrow linewidth transitions that have been used for imaging and cooling~\cite{norcia2018microscopic, cooper2018alkaline}, even ground state cooling in Ref.~\cite{Alexandre2018Alkaline}. Alkali atoms also contain narrow lines from forbidden transitions like the quadrupole transition of the S to D transition~\cite{saffman2016quantum} or (n)S to (n+1)P transitions~\cite{duarte2011all}. 

While sideband cooling with different trapping frequencies has been investigated through simulations~\cite{berto2021prospects}, in this section we derive a novel analytical formula for capturing the effects of different trapping frequencies, which also demonstrates additional opportunities for cooling on higher order sidebands.

We now derive the equations for an atom with a ground state frequency $\nu$ and an excited state frequency $\nu_e$.  The excited state harmonic eigenstates are stretched or compressed relative to the ground state eigenstates. The squeezing operator from quantum optics can therefore be utilized to relate the wavefunctions in the two potentials~\cite{taieb1994coolinga,martinez2018state,Urech2023Single}:
\begin{equation}
    |m_e \rangle = \hat{S}^\dag(r) | n_e \rangle,
\end{equation}
where $|n_e \rangle $ is the ground state basis  $|m_e \rangle$ is the excited state basis.
The squeezing operator $\hat{S}(r)$ is 
\begin{equation}
    \hat{S}(r) = \exp{ \left( \frac{r}{2}  (\hat{a}^2 - \hat{a}^{\dag 2})  \right)  }.
\end{equation}
The squeezing parameter $r$ is related to the trapping frequencies by
 \begin{equation}
    r = \frac{1}{2}\ln \left( \frac{\nu_e}{\nu} \right)  .   % from Urech 2023 Strontium thesis.  Agree numerically exactly with the Taieb expression, but simpler. Use this.    I took sqrt and - out of log. 
 \end{equation}

The excited state annihilation operator is related to the ground state annihilation operator by
\begin{equation}
    \hat{a}_e = \hat{S}^\dag(r) \, \hat{a} \, \hat{S}(r) .
\end{equation}
The harmonic oscillator Hamiltonian is
\begin{align}
    H_\text{ho} &=\nu ( \hat{a}^\dag \hat{a} )P_g + \nu_e( \hat{a}_e^\dag \hat{a}_e) P_e  \nonumber \\
    &= \nu  ( \hat{a}^\dag \hat{a} )P_g +  \nu_e \,\, \hat{S}^\dag(r)  (\hat{a}^\dag \hat{a} )\hat{S}(r) \,\, P_e.
\end{align}

The transfer rates between the ground states are similar to Eq.~\ref{eq:dopplerrate}, where now the squeezing operators are used to modify the excited harmonic states: 
\begin{gather} \label{eq:squeezingrate}
    \gamma_{n_1 \rightarrow n_2} =
     \Gamma  \,\, \Omega^2   \\ 
    \times \sum_{\scriptscriptstyle \pm}  \left|  \sum_{n_e} \frac{ \langle n_2 | e^{\pm i k \hat{x}} \hat{S}^\dag(r) | n_e  \rangle  \langle n_e| \hat{S}(r) e^{i k \hat{x}} | n_1 \rangle}{ ( \Delta + \nu n_1 - \nu_e n_e  ) - i \Gamma/2 }  \right|^2  . \nonumber
\end{gather}
 {Unlike the sideband cooling case with equal trapping frequencies,} the sum over the excited states cannot be taken out of the magnitude now because each term does not necessarily have the directional emission terms. 
There are additional terms now in the first order expansion due to the squeezing operator.  To first order in $r$ and $\eta$, the matrix elements are 
\begin{gather}
   \langle n_e | \hat{S}(r) e^{ik \hat{x}} | n_1 \rangle  \nonumber \\
  = \langle n_e | \left(  1 +  i \eta (\hat{a} + \hat{a}^\dag) +\frac{r}{2} (\hat{a}^2 -{(\hat{a}^\dag)}^2)\right)  | n_1 \rangle .
\end{gather}
The first order LD terms still couple $n \rightarrow n \pm 1$.  In contrast, the first-order squeezing terms couple $n \rightarrow n \pm 2$, preserving the  {even and odd} symmetry of the wavefunction.  

The resonance condition for the light absorption is now determined by the detuning $\Delta + \nu n_1 - \nu_e n_e $.  We can re-express this in terms of the difference trapping frequency $\Delta \nu = \nu_e - \nu$, which gives  $\Delta + \nu n_1 - \nu_e n_e = (\Delta +  n_e \Delta \nu) + \nu (n_1 - n_e)  $.
For increasing $n$, the detuning between $n_1$ and $n_e$ changes by $n_e \Delta \nu $.  We define the resonance condition for the $n$'th harmonic state as
\begin{equation}
    \Delta_n = \Delta +  n \, \Delta \nu.
\end{equation}

The $n \rightarrow n \pm 1$ transition are similar to before
\begin{gather}
    \gamma_{n \rightarrow n+1} =  \eta^2 \, (n+1) \, A_{\scriptscriptstyle +} \\
    \gamma_{n \rightarrow n-1} =  \eta^2 \, n \, A_{\scriptscriptstyle -} .  \nonumber
\end{gather}
The rates are a function of the $n$ dependent detunings 
\begin{gather}
    A_{\scriptscriptstyle +} =   R({\Delta}_n - \nu )   + \alpha R({\Delta}_{n} )  \\
    A_{\scriptscriptstyle -} =   R({\Delta}_{n} + \nu )   +  \alpha R({\Delta }_{n} ).  \nonumber
\end{gather}
In this context, the rates add incoherently due to the randomized phase acquired during emission in multiple directions.

As illustrated in Fig.~\ref{fig:Squeezing}(a), the $n$-dependent detuning results in the first-order cooling sideband being resonant only for a specific vibrational quantum number $n$.
For a less tightly trapped excited state (left side), the cooling light shifts towards heating sidebands for higher $n$ values. Conversely, for a more tightly trapping excited state (right side), lower $n$ values gradually shift towards cooling of higher-order sidebands. This effect leads to instability when $\nu_e/\nu < 1$, where higher-energy populations heat out of the trap. When $\nu_e/\nu > 1$, a type of ``cap" emerges, with heating occurring below a certain $n$ value and cooling above it. Consequently, this cap effectively traps population below it.

The squeezing operator now also allows   $n \rightarrow n \pm 2$ transitions to first order in $r$:
\begin{gather}
      \gamma_{n \rightarrow n+2} =   r^2 \, (n+2) (n+1)  \,\, B_{\scriptscriptstyle +} \\
    \gamma_{n \rightarrow n-2} =   r^2 \, n (n-1)    \,\, B_{\scriptscriptstyle -}  . \nonumber
\end{gather}
The new rates $B$ are a sum over different excited states, but the interference now matters because the coupling does not involve spontaneous emission to first order, 
\begin{gather}
     B_{\scriptscriptstyle -}= \left| {T}( {\Delta}_n) -  {T}({\Delta}_n  + 2 \nu  )       \right|^2 \\
     B_{\scriptscriptstyle +} = \left| {T}({\Delta}_n ) -  {T}({\Delta}_n  - 2 \nu )     \right|^2 .   \nonumber
\end{gather}
The complex scattering rates are
\begin{equation}
    T(\Delta) = \sqrt{\Gamma} \frac{\Omega/2}{ \Delta - i \Gamma/2}.
\end{equation}
and are related to the $R$ rates by  $|T(\Delta)|^2 = R(\Delta)$.  Interestingly, the destructive interference between the two excited pathways in the $n \rightarrow n \pm 2$ transition results in a transfer rate that cancels for large detuning, leading to much lower heating rates at large detuning. 

Next we estimate the final temperature as a function of $\Delta \nu$.  We construct a rate equation for $\langle n \rangle $  by multiplying the rate equation in Eq.~\ref{eq:rate} by $n$ and summing over all $n$.  This leads to the rate equation
\begin{gather}
    \frac{d \langle n \rangle}{dt} = 
\eta^2 \left[ A_{\scriptscriptstyle  +} (\langle n \rangle + 1) - A_{\scriptscriptstyle -} \langle n \rangle  \right] \nonumber \\
+ 2 r^2 \left[  B_{\scriptscriptstyle +} \langle (n+2)(n+1)\rangle - B_{\scriptscriptstyle -} \langle n (n-1)\rangle    \right]   .  
\end{gather}
Although we would need a separate expression for $\langle n^2 \rangle$ to solve this exactly, we can find an approximate solution by assuming the population follows a geometric series with a Boltzmann factor, as was the case in the Doppler section.  Then we can use the property $\langle n^2 \rangle = 2 \langle n \rangle^2 + \langle n \rangle$ for geometric distributions and solve the equation numerically to estimate the steady-state $\langle n \rangle$, which gives the equation
\begin{multline}
\label{eq:squeezinganalytical}
   0 =  \eta^2 \left[ \langle n \rangle (A_{\scriptscriptstyle  +} -A_{\scriptscriptstyle  -}  
  )  + A_{\scriptscriptstyle  +}  \right] \\
+ 4 r^2 \left[ \langle n \rangle^2 (B_{\scriptscriptstyle +}  -B_{\scriptscriptstyle -}  )   +  B_{\scriptscriptstyle +} (2 \langle n \rangle + 1) \right]  .   
\end{multline}
$B_\pm$ and $A_\pm$ are also $n$ dependent through the detuning $\Delta_n$.
This solution agrees well for small $r$ and $\eta$, with the main assumption that the distribution follows a geometric series. This will break down for high populations as the cooling and heating become $n$ dependent.

The steady-state population is a balance between the sideband and squeezing terms.  {Interestingly, the squeezing does allow additional cooling by making the $n-2$ sideband larger than the $n+2$ sideband.} Cooling due to sidebands is $ \eta^2 (A_- - A_+)$, while squeezing cooling is $2 r^2 (B_- - B_+)$.  { The squeezing dependence is quadratic with $n$ and is more important at larger $n$.} Additionally, there is heating from both that goes as $\eta^2 A_+$ and $2r^2 B_+$, respectively.   {This new coupling does however bring up an interesting question about whether it can provide new opportunities for cooling. }

Fig.~\ref{fig:Squeezing}(b) is a master equation simulation of steady-state population versus the excited to ground state trapping frequency ratio $\nu_e/\nu$. The magic condition is $\nu_e/\nu=1$.   Cooling performance drops on both sides, with much more dramatic for a weaker excited state ($\nu_e/\nu < 1$).  
 Fig.~\ref{fig:Squeezing}(c) is a detuning scan for $\eta=0.1$, $\nu = 2 \Gamma$,  and various $\nu_e/\nu$.  There is a resonance at $\Delta = - \nu$ due to sideband cooling, but also a second resonance at $\Delta = -2\nu$ due to the squeezing coupling.    

For cooling the atom in a `non-magic' trap, a technique known as ``Sisyphus Cooling", has been recently demonstrated in tweezers~\cite{cooper2018alkaline, urech2022narrow}. In situations where the polarizability of the excited state is greater (or less) than that of the ground state, the atom experiences a deeper (or shallower) trap upon excitation by the cooling laser. This difference enables the implementation of `attractive' (or `repulsive') Sisyphus cooling. In the repulsive Sisyphus cooling method, there is a limit known as the Sisyphus cap to how much the atom can be cooled. However, in the attractive Sisyphus regime, significantly lower temperatures can be achieved within the trap. Ref.~\cite{berto2021prospects} proposes that sweeping the cooling laser's frequency adiabatically could achieve temperatures well below the Sisyphus cap.

%%%%%%%%%%%%%%%%%%%%%%%%%%%%%%%%%%%%%%%%%%%%%%%%%%%%%%%%%%
\section{Sub-doppler cooling with spin} \label{sec:spin}
\begin{figure*}[bt!]
  \centering
  \includegraphics[width=0.9\linewidth]{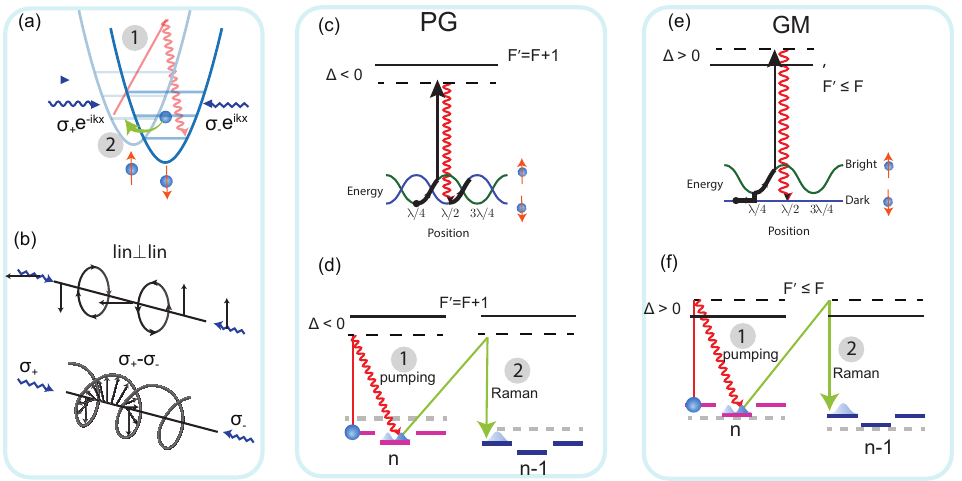}
  \caption{ PG and GM spin cooling. . \textbf{(a)} An atom with spin in a harmonic trap with changing polarization.   {Counter-propagating light with different polarizations couple different motional states with different spins. PG and GM utilize the light for optical pumping, Raman transfer, and the light shifts to make the Raman transfer resonant.} \textbf{(b)} Lin$\perp$lin and $\sigma_+-\sigma_-$ polarization configurations.   {Lin$\perp$lin changes from linear to circular, whereas $\spsm$ is a rotating linear field. } \textbf{(c-d)} Free-space (top) and trapped picture (bottom) for PG cooling, which uses $F \rightarrow F+1$ and red detuning. Population is optically pumped into the lower energy state. \textbf{(e-f)} Free-space (top) and trapped picture (bottom) for GM cooling, which uses $F' \le F$ and blue-detuning to create dark states, which act as the lower energy state into which the population is pumped.}
  \label{fig:spinintro}
\end{figure*}
Doppler or sideband cooling alone is often insufficient for achieving ground state cooling of neutral atoms in optical traps due to two main issues. First, the trapping frequencies in optical traps are typically less than 100 kHz, while the line widths of the D1 and D2 transitions are around 5 MHz, placing them far into the unresolved regime. Second, the ground and excited states have different optical polarizabilities, which can lead to additional heating and atom loss.

To overcome these limitations, sub-Doppler cooling schemes such as polarization gradient (PG)~\cite{dalibard1989laser}, gray molasses (GM) and $\Lambda$-enhanced GM~\cite{salomon2013gray, grier2013lambda}, electromagnetically induced transparency (EIT) cooling~\cite{Giovanna2000Ground},  and Raman sideband (RSB)~\cite{monroe1995resolved} employ two-photon transitions between ground states, as illustrated in Fig.~\ref{fig:spinintro}(a). By detuning far from the excited states and not populating them,  these techniques mitigate the issues associated with the unresolved regime and differential polarizabilities. Furthermore, different ground hyperfine states have the same scalar light shift and no tensor shift, resulting in the same trapping frequencies for different spin states in linear trapping light.

Many of these schemes were developed in the free-space picture, where the atom transverses many wavelengths.  In this section, we present a unified spin cooling formalism for trapped atoms that encompasses PG, GM, EIT, $\Lambda$-GM, and RSB cooling.  They are all two-photon cooling schemes that rely on coupling two ground spin states with different motional quantum numbers and energies.  {For example, the free-space and bound pictures for PG and GM are shown in Fig.~\ref{fig:spinintro}(c-f).}

\del{The quantum model for PG and GM cooling is depicted in Fig.~\ref{fig:spinintro}(d,f).  In the first stage, a two-photon coherent transition (green) changes the spin and motional states, where the motional state decreases from $n$ to $n-1$. In the second stage, spontaneous emission (red) from the cooling light or a separate beam then re-pumps the atom back to the original spin state, allowing the cooling cycle to repeat.  As long as the trapping frequency energy $\hbar \nu$ is larger than the recoil energy, net cooling can be achieved.}

We show that all these schemes exhibit similar forms of the effective Hamiltonian and collapse operators in the LD regime and rely on the same principles. 
PG and GM cooling use a single laser frequency counter-propagating with a different polarization to create a polarization gradient. The cooling is optimal when the light shifts are similar to the trapping frequency.  This same laser is also used to generate the Raman coupling and optical pumping. EIT and $\Lambda$-GM cooling incorporate a second coherent laser resonant to a different ground spin state to create dark states of a three-level $\Lambda$ system.  This improves cooling by creating darker reservoir states.  RSB cooling separates all three processes, using two lasers and their relative detuning to couple different ground spin states and a third frequency-beam for optical pumping that uses angular momentum selection rules to keep the cooling state dark.

\subsection{Polarization Gradient Cooling and Gray Molasses} \label{subsec:polarization}
PG cooling was described theoretically by Cohen-Tannoudji in the late 1980's~\cite{dalibard1989laser} and resulted in sub-doppler temperatures in the first magneto-optical traps~\cite{Lett1988Observation}.  Two different PG models were presented in free space for the case of lin$\perp$lin and $\spsm$  polarizations, both shown in Fig.~\ref{fig:spinintro}(b).  
Both forms of PG cooling work by using red-detuned light with a $F \rightarrow F+1$ transition, which is the same configuration used by a MOT for its stretch state cycling condition. 

The lin$\perp$lin configuration comes from counter-propagating orthogonal linear polarization beams, resulting in a polarization that changes from horizontal to right-hand circular to vertical to left-hand circular every wavelength. Because it relies on a vector shift rather than a tensor shift, the simplest spin structure that this model works on is $F=1/2$ to $F'=3/2$ atom. The free-space picture is shown in Fig.~\ref{fig:spinintro}(c).  The circular polarization results in a vector shift $\Delta E \propto m_F$ that modulates every wavelength.  PG cooling works by having an optical pumping that pumps the atom to the lower potential.  As the atom travels, it goes up the potential hill, loses energy, and then is pumped back to the bottom.  The cycle repeats until the atom approaches the recoil temperature $E_\text{recoil}$, and can reach a $\mu$K. This style of cooling is also called Sisyphus cooling, named after the Greek mythological figure who was condemned to eternally push a boulder up a hill. 
%In a more classical picture, a polarizable atom moves through a changing polarization to which it is trying to align.  There is a lag due to the time $1/\Gamma$  it takes to orient with the field that results in a drag on the atom $F \propto -v/\Gamma$.
% The lag is proportional to both the atom's velocity $v$ and the time scale it takes for the atom to align with the changing field, which is the inverse of the optical pumping rate $\gamma$. This relationship produces a damping force.

Counter-propagating laser beams with opposite circular polarizations is called the $\spsm$ model.  As shown in the bottom of Fig.~\ref{fig:spinintro}(b), the resulting polarization is linear at every point in space but rotates along the propagation direction with a period equal to the wavelength.  Because there is no circular light and the resulting vector shift, the $\spsm$ model relies on a tensor shift $\Delta E \propto m_F^2$ and therefore requires a ground state spin of at least $F=1$.  The $\spsm$ model is much more common experimentally with neutral atoms because it is also the polarization configuration used by a magneto-optical trap (MOT).  

 {Interestingly, in free-space the $\spsm$ and lin$\perp$lin schemes have different pictures  and even inverse dependencies on detuning and intensity~\cite{dalibard1989laser}, however, they result in similar minimum temperatures.  The  $\spsm$ picture is not typically used because it is conceptually more difficult than the modulated vector shift picture of lin$\perp$lin.}

Gray Molasses (GM) is a closely related cooling scheme that operates with blue-detuned light with a $F \rightarrow F$ or $F \rightarrow F-1$ transition.  GM has achieved even colder temperatures than PG cooling~\cite{fernandes2012sub, salomon2013gray, nath2013quantum} by storing cold atoms in dark states that do not interact with the cooling light due to polarization and angular momentum selection rules, as shown in Fig.~\ref{fig:spinintro}(e).  Certain ground states are not coupled to the excited states.  GM works with blue detuned light so that the bright states are higher in energy.  Atoms that fall into these dark states are effectively shelved, reducing their interaction with the cooling light. The residual velocity couples the atom back into a bright state, where they can undergo Sisyphus cooling again.

However, these free-space models break down for a trapped atom in the Lamb-Dicke (LD) regime, where the atom's wavefunction is smaller than the cooling wavelength.  While PG and GM for bound atoms have been implemented for atoms in optical lattices {~\cite{Winoto1999Laser}}, tweezers~\cite{blodgett2023imaging, darquie2005controlled, huang2022gray}, and ions~\cite{joshi2020polarization, li2022robust}, the mechanism has not been described theoretically.  The most related works have been Wineland et al. in Ref.~\cite{wineland1992sisyphus}, which studied a bound atom trapped in a linearly polarized standing wave, and Cirac et al. in Ref.~\cite{cirac1992laser} which investigated a $J=1/2 \rightarrow J'=3/2$ atom in the lin$\perp$lin configuration. Both of these studies found a lower limit for the mean vibrational quantum number, $\langle n \rangle \approx 1$. Here we develop a generalized spin cooling model that applies to all spins and polarizations and shows that certain configurations can lead to high fidelity ground state cooling. 

\subsection{Spin cooling theory}  \label{subsec:spin}
%We start with the atom-field master equation from Eqs.~\ref{eq:Hdetuned, eq:Ldetuned}.
Here we develop the spin cooling master equation, including both the ground and excited states, for PG and GM.  
First we look at the case of two counter-propagating beams with opposite-handed circular polarized light (\spsm).  
For an atom moving along $z$, the complex polarization is 
\begin{gather} \label{eq:spsm}
\bm{\epsilon}_{\spsm}(\hat{z}) =  \frac{i}{\sqrt{2}} \left( e^{i k \hat{z}} \hat{\bm{e}}_{1}  +  e^{- i k \hat{z}} \hat{\bm{e}}_{-1} \right) \nonumber  \\
 = \sin{(k \hat{z})} \, \hat{\bm{x}} +  \cos{(k \hat{z})} \, \hat{\bm{y}}  \nonumber \\
 \approx  \bm{\hat{y}} + (k \hat{z}) \, \bm{\hat{x}}
\end{gather}
where $\hat{\bm{e}}_{1,0,-1}$ are the spherical unit vectors $\hat{\bm{e}}_{\pm1} = \mp \frac{1}{\sqrt{2}} (\hat{\bm{e}}_x \pm i \hat{\bm{e}}_y)$ and $\hat{\bm{e}}_0 = \hat{\bm{e}}_z$. The field is Taylor expanded in the last line.   {We can see from this expression that the atom experiences a rotating linear field, as shown in Fig.~\ref{fig:spinintro}(b).}  The intensity is constant everywhere. In the LD regime, As it oscillates back and forth in the trap, it sees a small rotation of the field. For convenience, we now choose a different coordinate system. We choose the quantization axis $\hat{\bm{z}}$ to be aligned with linear polarization at equilibrium.  The cooling is however independent of the atom position because the light is linear everywhere.

Substituting the polarization into Eqs.~\ref{eq:Hdetuned} and \ref{eq:Ldetuned}, the Hamiltonian and collapse operators for spin cooling (PG and GM) for $\spsm$ in the LD regime are
\begin{gather}   \label{eq:Hsigma}
H_{\sigma_{\scriptscriptstyle +}-\sigma_{\scriptscriptstyle -}} =\nu \hat{a}^\dag \hat{a} + (\Delta - i \Gamma/2) P_e    \nonumber \\
+  \frac{\Omega}{\sqrt{2}}  \left[  \hat{D}^\dag_z - i k \hat{x} \hat{D}_x^\dag + h.c.   \right] \\
 \hat{L}_{q,\pm} = \sqrt{\Gamma} \, (1 \pm i k \hat{x}) \hat{D}_q 
\end{gather}
The Cartesian dipole operators are defined as  $\hat{D}_i = \hat{\bm{i}} \cdot \hat{\bm{D}}$.   { For example, the $\hat{D}_z$ operator raises all of the ground state spins to the excited state with the same spin $m_{F'} = m_F$.  The operator $\hat{D}_x = (\hat{D}_{1} + \hat{D}_{-1})/\sqrt{2}$ creates a superposition of $m_{F'} = m_F \pm 1$.     $\hat{D}_q$ for $q = 0, \pm 1$ still represents the spherical vector basis.}

 {
This master equation both represents PG and GM cooling.  For example,  Fig.~\ref{fig:spinintro}(d,f) show the trapped PG and GM processes for an $F=1$ atom.  In the trapped situation, the picture for both cases are similar.  PG is the case of $F'=2$ while is GM is the case of $F'=1$. The Clebsch-Gordon for these two cases are also shown in Fig.~\ref{fig:pg}(a) and (c). The position-independent interaction $\hat{D}_z$ leads to a tensor shift and optical pumping.  In both cases, the population pumps to the $m=0$ state.  However, in the PG case the $m=0$ state is bright, whereas in the GM case it is dark.  In order to make the $m=0$ state the lower energy, the PG case has to work with red detuning, while the GM case has to work with blue detuning. This is first ingredient for spin cooling: the population is pumped to the lower energy spin state. 
}

 {
The second ingredient for spin cooling is a two-photon Raman transition between the $|m=0, n \rangle$ and $|m=\pm1, n-1 \rangle$, shown in green.  This two photon transition arises from the position independent term and the position dependent terms in the interaction.  
The position-dependent term in the interaction $ \hat{x} \hat{D}_x$ drives an orthogonal polarization transition $m \rightarrow m \pm 1$ while also changing the motional state.   As shown in the illustration, together these terms produce a coherent transition from $|m=0, n \rangle$ to $|m=\pm1, n-1 \rangle$. As drawn, the resonance for this two-photon condition is satisfied when the differential light shift between the $m=0$ and $m=\pm1$ states are equal to the trapping frequency.  
}

 {
This process then repeats, cooling the atom until the cooling and recoil heating from spontaneous emission reach an equilibrium. As we will see later in the full simulations, GM is much better at preparing ground state population due to the presence of the dark state.  
}

This expansion can be performed for other polarizations as well to get a similar form.  We can also look at the lin$\perp$lin case, where the polarization changes from circular to linear depending on the atom position,
\begin{equation}
     \bm{\epsilon}_{\text{lin} \perp \text{lin}}(\hat{x}, \phi) =  \frac{1}{2} ( e^{i k \hat{x}} \hat{\bm{x}}  + e^{i \phi} e^{- i k \hat{x}} \hat{\bm{y}} )  . 
\end{equation}
 {In this case, however, the position of the atom plays a much more important role because it changes the polarization at equilibrium position}  For circular polarization, the interaction is 
\begin{equation}
H^I_{\text{lin} \perp \text{lin}}(\phi=\pi/2) = \frac{\Omega}{2} \left[ \hat{D}_{1}^\dag + k \hat{x} \hat{D}_{-1}^\dag + h.c. \right],
\end{equation} 
and for linear polarization, 
\begin{equation}
H^I_{\text{lin} \perp \text{lin}}(\phi=0) = \frac{\Omega}{2} \left[ \hat{D}_{z}^\dag +k \hat{x} \hat{D}_{x}^\dag + h.c. \right]. % check sign here
\end{equation} 

In the trapped model, the distinction between PG, GM, $\spsm$, and lin$\perp$lin breaks down, and we can refer to them more generally as spin cooling. 

\subsubsection*{Effective spin ground state operators}
The cooling and heating processes in the spin model can be seen more clearly by looking at the effective ground state operators with the excited states adiabatically eliminated.  {Because PG and GM cooling typically use a detuning of ten linewidths or more away from the excited resonance, this is typically a very good approximation.}  From Eq.~\ref{eq:Hdetuned}, the effective ground state Hamiltonian for the $\spsm$ case is
\begin{gather}
H^\text{eff}_{\sigma_{\scriptscriptstyle +}-\sigma_{\scriptscriptstyle -}} = \nu \hat{a}^\dag \hat{a} - \\
\frac{\Omega^2}{4 \Delta } \left[ \hat{D}_z \hat{D}_z^\dag + k \hat{x} ( \hat{D}_z \hat{D}_x^\dag + \hat{D}_x\hat{D}_z^\dag ) \right] \nonumber . 
\end{gather}
Converting it to spin operators using Eq.~\ref{eq:Hdetuned2} gives an effective spin model
\begin{gather} 
    H^\text{eff}_{\sigma_{\scriptscriptstyle +}-\sigma_{\scriptscriptstyle -}} = \nu \hat{a}^\dag \hat{a} +  \frac{\Omega^2}{\Delta } \left[ C^{(0)} + C^{(2)} (\hat{F}_z^2 -  \hat{F}^2/3 )\right]  \nonumber  \\
    +  k\hat{x} \,\, \frac{\Omega^2}{\Delta } \, C^{(2)} \, (\hat{F}_x \hat{F}_z + \hat{F}_z \hat{F}_x )   .\label{eq:HAspsm}
\end{gather}
The collapse operators are 
\begin{gather}   \label{eq:LAspsm}
    L_{q, \pm}^{\text{eff}} 
    =\sqrt{\Gamma} \frac{\Omega/2}{\Delta - i \Gamma/2}  \left[  \hat{D}_q \,  (\hat{D}_z^\dag + i k\hat{x}(\hat{D}_z^\dag+  \hat{D}_x^\dag) ))\right]  .
\end{gather}
A spin version of this operator can also be found using Eq.~\ref{eq:Ldetuned2}, although it is not written here. 
These operators are valid for all ground and excited spin combinations. While the ground state spin determines the spin operators $F_i$, the excited state spin only affects the $C^{(i)}$ coefficients in the spin representation.  

The Hamiltonian and collapse operator both have a position-independent and position-dependent term.   {As a result, the interpretation becomes similar to the sideband cooling case, where the position independent term drives $n \rightarrow n$, and the position dependent term drives $n \rightarrow n\pm1$.  However, now the operators act within the ground state spin manifold rather than the between the ground and excited states.}

\subsubsection*{Simplified spin model}
\begin{figure}[h!]
  \centering
  \includegraphics[width=0.7\linewidth]{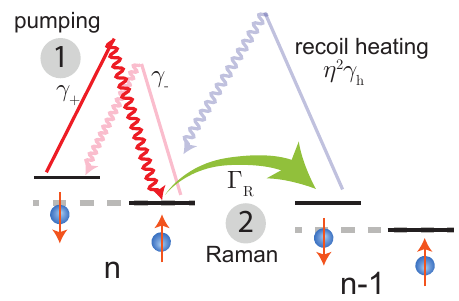}
  \caption{ {Simplified spin model with spins $\uparrow$ and $\downarrow$. The motional states $n$ and $n-1$ are separated by trapping frequency $\nu$.  The polarization shifts $| \uparrow,n \rangle$ into resonance with $|\downarrow ,n-1\rangle$.  The light optical pumps the two states between spins with rates $\gamma_+$  and $\gamma_-$.  The two-photon Raman transition is driven at rate $\Gamma_R$.  Photon recoil also heats the atom from $n$ to $n+1$ at rate  $\gamma_h$.} }
  \label{fig:spinmodel}
\end{figure}

 {The ground state spin models accurately predict population transfer, however, they are difficult to use for analytical models.  Here we develop a spin 1/2 model, from which we derive simple analytical expressions that we use to understand the final temperature.  The spin model is shown in Fig.~\ref{fig:spinmodel}. }

\begin{figure*}[tb]
  \centering
  \includegraphics[width=0.99\linewidth]{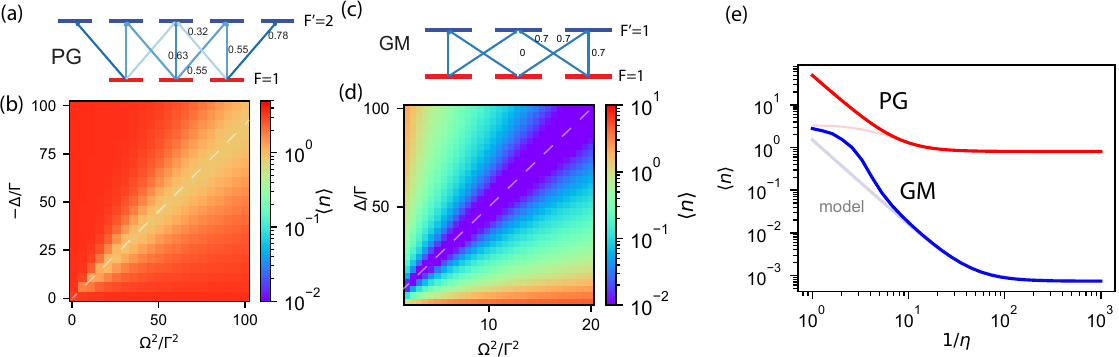}
  \caption{ PG and GM simulations.  \textbf{(a-b)} Clebsch-Gordon coefficients for PG ($F=1$ to $F'=1$) and GM ($F=1$ to $F'=2$). For PG in linear light, the population pumps to the $m=0$ state, which is also the brightest state.  PG cooling requires red-detuned light $(\Delta < 0)$. For GM in linear light, the population pumps to $m=0$ which is dark. GM cooling requires blue-detuned light $(\Delta>0)$.   \textbf{(c-d)} Simulations of $\langle n \rangle$ versus cooling power for PG and GM.  Optimal cooling occurs on a line (light-dashed) where the differential light shift equals the trapping frequency.  PG is limited to $\langle n \rangle  \approx 1$, while GM cools close to the ground state due to the dark state.  Simulations use $\eta =0.05$ and 8 harmonic states.  \textbf{(e)}  Optimal cooling for PG and GM versus the Lamb-Dicke parameter $\eta$. Optimal cooling occurs when the differential light shift is equal to the trapping frequency and in the limit of large detuning.   The simple model from Eq.~\ref{eq:simplemodel} is plotted in a lighter color. For PG, $\Delta = -20 \Gamma$ and for GM $\Delta = 20 \Gamma.$ The optical cooling powers were adjusted to make the Raman transition resonant.  }
  \label{fig:pg}
\end{figure*}

We assume two spin states $|\!\!\down,n \rangle$ and $|\!\!\up, n \rangle$.  We use an effective Hamiltonian and collapse operators that are similar in form to Eqs.~\ref{eq:HAspsm} and \ref{eq:LAspsm}.  We represent the light shift with $V_0$ and Raman coupling with  $\Omega_R$ in the Hamiltonian and collapse operators
\begin{gather}      
    H = \nu \hat{a}^\dag \hat{a} + V_0( \hat{F}_z + 1/2) + \Omega_R  k \hat{x} \,  \hat{F}_x   \nonumber \\
    L_{1} =\sqrt{\gamma_{\scriptscriptstyle +}}\,\, (\hat{F}_{+}  + i k \hat{x} \hat{F}_{x} )  \nonumber \\
    L_{-1} = \sqrt{\gamma_{\scriptscriptstyle -}}\,\, (\hat{F}_{-}  + i k \hat{x} \hat{F}_{x} )  . 
\end{gather}
 {The light shift is determined by $V_0$ and is defined such that $m_F=-1/2$ or $|\downarrow \rangle$ has zero shift.  The Hamiltonian also contains a position dependent term $\Omega_R k \hat{x} \hat{F}_x$.  This term produces Raman transitions between $|\up,n\rangle$ and $|\down,n-1\rangle$.  The two collapse operators also contain a position independent and position dependent part.  The position independent part produces optical pumping within the same $n$.  The pumping rate from $| \downarrow \rangle$ to $| \uparrow \rangle$ is $\gamma_+$, and the inverse optical pumping $\gamma_-$, as shown in Fig.~\ref{fig:spinmodel}. }

Optimal cooling occurs when the light shift is equal to the trapping frequency $\nu$, so that $|\up, n \rangle$ and $|\down, n-1 \rangle$ are the same energy. Then the Raman transition is resonant.   The collapse operators also contain a position dependent term that couples different motional states and represents recoil heating. 
% The $|\down, n \rangle$ couples to $|\up, n-1 \rangle$, but off-resonantly so we can ignore it.

 {Now we derive an approximate expression for the steady-state temperature of this model.  First we estimate the cooling rate due to the Raman transition.  The position dependent Hamiltonian coherently couples $|\up, n \rangle$ to $|\down, n-1 \rangle$. However, this alone would not result in cooling because the population would Rabi oscillate back and forth.  The optical pumping rates $\gamma_+$ and $\gamma_-$ produce decoherence that results in population transfer.
We can estimate this transfer rate by first solving for the steady state coherence between the Raman-coupled states using the master equation for the density matrix and setting the time derivative to zero:}
\begin{gather}
\langle \up, n |\rho|\down, n \! - \! 1 \rangle =\
\frac{i  \sqrt{n}\,  \eta \,\Omega_R }{ \Delta -i (\gamma_{\scriptscriptstyle -} \!+\!\gamma_{\scriptscriptstyle +})/2 }  
 \,\left[ P_{\scriptscriptstyle \up}(n) - P_{\scriptscriptstyle \down}(n\!-\!1)  \right]  
\end{gather}
By substituting this coherence into the density matrix equations and ignoring off-resonance coherences, we obtain the transfer rate $\Gamma_R$ between $| \uparrow, n \rangle $ to $|$ $\downarrow, n-1 \rangle$:
\begin{equation} \label{eq:GammaR}
\Gamma_R = \frac{4 \eta^2 \Omega_R^2 }{(\gamma_{\scriptscriptstyle +} + \gamma_{\scriptscriptstyle -})} \,\, \frac{1}{1 + 4 \Delta_{r}^2 / (\gamma_{\scriptscriptstyle +} + \gamma_{\scriptscriptstyle -})^2}.
\end{equation}
$\Delta_r$ is the relative detuning between the two Raman coupled states.  Optimal cooling occurs when the light shift matches the trapping frequency and brings these states into resonance, i.e., $\Delta_r = 0$. Although there are off-resonant transitions, such as $| \down,n\rangle \rightarrow |\uparrow ,n-1\rangle$, we ignore them. By neglecting all off-resonant coherences, we arrive at the population rate equations:
\begin{gather} \label{eq:PratePGGM}
\frac{dP_{\scriptscriptstyle \down}(n)}{dt} = \Gamma_{R} \,\, (n\!+\!1) \left[ -P_{\scriptscriptstyle \down}(n) \!+\! P_{\scriptscriptstyle \up}(n\! +\! 1)\right]    \\
+ \gamma_{\scriptscriptstyle -} P_{\scriptscriptstyle \up}(n) - \gamma_{\scriptscriptstyle +} P_{\scriptscriptstyle \down}(n)   \nonumber \\
+\eta^2 \gamma_{\scriptscriptstyle h} \left[ (n+1)P_{\scriptscriptstyle \up}(n\!+\!1) + n P_{\scriptscriptstyle \up}(n\!-\!1)  - (2n+1) P_{\scriptscriptstyle \down}(n)\right]  \nonumber  \\
\frac{dP_{\scriptscriptstyle \up}(n)}{dt} =n  \Gamma_R \, \left[P_{\scriptscriptstyle \down}(n-1) - P_{\scriptscriptstyle \up}(n)\right]  \nonumber   \\
- \gamma_{\scriptscriptstyle -} P_{\scriptscriptstyle \up}(n) + \gamma_{\scriptscriptstyle +} P_{\scriptscriptstyle \down}(n)  \nonumber \\
+ \eta^2 \gamma_{\scriptscriptstyle h} \left[ (n+1)P_{\scriptscriptstyle \down}(n\! + \! 1) + n P_{\scriptscriptstyle \down}(n\!-\!1)  - (2n+1) P_{\scriptscriptstyle \up}(n)\right]    \nonumber
\end{gather}
We have defined a  total spontaneous emission rate  \[\gamma_h = \gamma_{+} + \gamma_{-}.\]

Next, we calculate the steady state solution for $\langle n \rangle$.  The derivation is as follows.  First, we perform the sum $\sum_{n=0}^\infty$ to the two equations in Eq.~\ref{eq:PratePGGM}.  Next we multiply these two equations by $n$ and perform the same sum, yielding two more equations.  This results in a total of four equations.  We set the time derivatives to zero to calculate the steady-state.  We then have six unknowns in these four equations:  $\langle n^2 \rangle_\up$, $\langle n^2 \rangle_\down$,   $\langle n \rangle_\up$,  $\langle n \rangle_\down$,  and the fraction of population in either up or down 
\begin{align}
    P_\up = \sum_{n=0} P_\up(n), \quad  P_\down = \sum_{n=0} P_\down(n).
\end{align}
Two of these terms are higher-order moments.  If we assume that $P_\uparrow(n)$ and $P_\downarrow(n)$ follow a geometric distribution, then we can use properties of the geometric series to write them as
\begin{gather}
    \langle n^2 \rangle_{\up} =  2 \langle n \rangle_{\up} / P_{\up} + \langle n \rangle_{\up} \\\langle n^2 \rangle_{\down} =  2 \langle n \rangle_{\down} / P_{\down} + \langle n \rangle_{\down}.
\end{gather}
Here  $P_\up$ and  $P_\down$ appear because $P_\up(n)$ and $P_\down(n)$ are not a normalized distribution as they do not sum up to one.  

 {
We then have four equations and four unknowns.  Then we solve the system of equations for  $\langle n \rangle_\up + \langle n \rangle_\down = \langle n \rangle$.  This expression is long, but we can find a simpler expression by taking the limit of large detuning $\Delta$. There are two main processes in the time evolution.  The Raman transfer $\Gamma_R$ and the pumping rates $\gamma_\pm$.   In the large detuning limit, the cooling rate  
 $\Gamma_R = 4 \Omega_R^2 \eta^2 / (\gamma_{\scriptscriptstyle +} + \gamma_{\scriptscriptstyle -})$ dominates.  The pumping rates scale as $1/\Delta^2$.  
 The cooling rate $\Gamma_R$ is independent of detuning because both $\Omega_R^{2}$ and the scattering rate scale as $1/\Delta^{2}$.  Consequently, for large detuning, $\Gamma_R$ becomes the dominant rate, and the population quickly distributes between the two Raman-coupled states. The asymmetry of the optical pumping then determines the cooling efficiency and the final temperature.
 }
 
In this limit of large detuning, the Raman coupling rate is faster than any scattering rate, and the steady-state population simplifies to
\begin{equation} \label{eq:simplemodel}  
     \langle n \rangle = \frac{s}{1-s}, \quad
s =  \frac{ 4 \eta^2 \gamma_h  + \gamma_{\scriptscriptstyle -} }{\eta^2 \gamma_h + \gamma_{\scriptscriptstyle +} } .
\end{equation}
 {This expression for the steady-state population is the final result of this model. It agrees very well with the exact spin 1/2 solution for $\eta < 0.1$.  We can look at a few limiting cases.  The ratio $\gamma_-/ \gamma_+$ is the pumping asymmetry between the spin states. For PG cooling this ratio is approximately one, as there is a small asymmetry.  The factor $\eta^2 \gamma_h$ is a spontaneous emission rate that leads to heating.  For PG and small  $\eta$, the steady-state population becomes $s \approx \gamma_-/\gamma_+$, which gives $\langle n \rangle \approx 1$.  And this does not depend on the LD parameter. }

 {Experimentally, both PG and GM operate in a low B-field regime. Since they use light shifts to adjust the motional trap frequencies into resonance, the Zeeman shifts must remain smaller than the trapping frequency, providing an estimate for how small the B-field needs to be.}

 {For the GM case though, the lower energy state is dark.  For $\gamma_- = 0$, then $ \langle n \rangle \approx 4 \eta^2 \frac{\gamma_h}{\gamma_+}$.  Because $\gamma_h = \gamma_+ + \gamma_-$, this ratio is close to unity.  This scales with with the LD parameter squared and can be made much smaller than in the PG case. }

 {
In the derivation of the steady-state temperature for the simplified model, we ignore the off-resonant coherent coupling between \( |\uparrow, n-1 \rangle \) and \( |\downarrow, n \rangle \). From Eq.~\ref{eq:GammaR} and using the off-resonant detuning \( \Delta_r = 2 \nu \), we obtain a transfer rate between these states of 
\[
\Gamma_{R}^\text{off} \approx \frac{ \eta^2 \Omega_R^2 \gamma_h}{ 4 \nu^2 }.
\]
The resonance condition is close to \( \Omega_R \approx \nu \), so we can eliminate those terms and get an approximate heating rate from off-resonant scattering:
\[
\Gamma_{R}^\text{off} \approx \frac{ \eta^2 \gamma_h}{ 4 }.
\]
This heating rate is four times less than the heating rate \( \eta^2 \gamma_h \), which arises from the position-dependent collapse operator in this model, and can thus be ignored. In the effective model, we can consider this off-resonant coherent transfer as being incorporated into the effective \( \gamma_h \).
}

 {In the next section we perform full simulations and compare it to the model.}  For other $F$, this model will also serve as a useful fitting function.  The ratio $\gamma_-/\gamma_+$ is then an effective asymmetry of the optical pumping to the lower energy state.  This asymmetry is close to unity for PG cooling, and much smaller for cooling with dark state, i.e. GM and $\Lambda$-GM.  The ratio $\gamma_h/\gamma_+$ is an effective scattering rate of the lower energy state.   Together with the Lamb-Dicke parameter $\eta$, these three parameters determine the lowest achievable population.  

%We note that a similar expression can also be derived from the no-flow condition solving for $s = P(n+1)/P(n)$ without having to take the expectation values, although other approximations have to be made. 

\subsubsection*{Simulations}
In Fig.~\ref{fig:pg}, we simulate the full master equation with ground and excited states for PG and GM for $F=1$ ground state in $\spsm$ light, meaning that the light is linear at the atom equilibrium position.  The PG case is $F=1$ to $F'=1$, and the Clebsch-Gordan (CG) coefficients are shown in (a).  In the \spsm configuration, the $m=0$ state has the largest CG and, consequently, the largest Stark shift.  In linear light, however, the population still pumps to the $m=0$ state because of strong diagonal CG coefficients from $|J'=2, m=\pm1 \rangle$.   Cooling requires that the population is pumped to the lowest energy state.  Therefore, cooling requires the red-detuned configuration where the $m=0$, which is the most stark shifted state, is also the lowest energy, as shown in Fig.~\ref{fig:spinintro}(d).    

In the PG simulation in Fig.~\ref{fig:pg}(b), cooling is optimum along the line where the Raman transition is resonant.  The optical power of the cooling light is such that the energy difference between $m=0$ and $m=\pm1$ equals the trapping frequency.  The lowest population converges to $\langle n \rangle = 0.9$ in the limit of large detuning $\Delta$ and small LD parameter $\eta$.  The optimal value is plotted in red in Fig.~\ref{fig:pg}(e). The light red line is the simple model from Eq.~\ref{eq:simplemodel} with pumping asymmetry $\gamma_-/\gamma_+ = 0.45 $, and heating scattering $\gamma_h/\gamma_+= 27$.  There is not a large asymmetry in the optical pumping, and $\gamma_{\scriptscriptstyle +} \sim \gamma_{\scriptscriptstyle -}$, which are both larger than $\eta^2 \gamma_s$ in the LD regime. In this case, the final population is
$s \approx \gamma_{\scriptscriptstyle +} / \gamma_{\scriptscriptstyle -}$.  In this situation, the cooling reaches a fundamental limit for large detuning and small $\eta$.  

Next we look at GM for $F=1$ to $F'=1$.  The CG coefficients are in Fig.~\ref{fig:pg}(c).  One optical selection rule is that the $|F, m=0 \rangle$ to $|F', m=0 \rangle$ is forbidden, due to the photon needing to add one unit of angular momentum.  Because of this dark state, all the population pumps to $m=0$.  But in contrast to PG, this state is now dark.  For that reason GM requires blue-detuned light ($\Delta>0$) to make this state the lowest energy.  The simulation in Fig.~\ref{fig:pg}(d) again shows that optimal cooling occurs when the light shift equals the trapping frequency. The population however drops close to ground state.   The optimal cooling is plotted in  Fig.~\ref{fig:pg}(e), along with the simple model with effective parameters $\gamma_-/\gamma_+ = 7.5 \times 10^{-4}$ and heating scattering $\gamma_h/\gamma_+= 1.5$. For this large optical pumping asymmetry where the reservoir state is dark, such as found in GM and EIT cooling, $\gamma_{\scriptscriptstyle -} = 0 $, and $s \approx \frac{\gamma_{\scriptscriptstyle +}}{\eta^2 \gamma_h} $.   In this dark cooling limit,  the final population is instead limited by $\eta$.

\begin{figure}[tb!]
  \centering
   \includegraphics[width=0.99\linewidth]{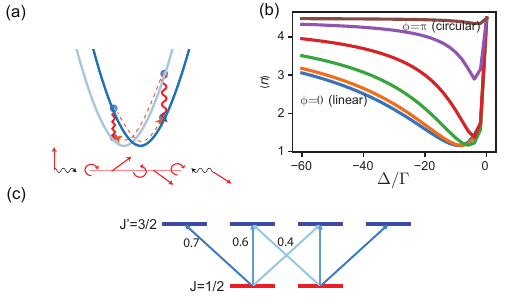}
  \caption{ $J=1/2$ to $J'=3/2$ PG cooling in lin$\perp$lin light. \textbf{(a)} Classical picture of cooling with spins in polarization gradient of  lin$\perp$lin light, inspired from Ref.~\cite{cirac1993laser}. \textbf{(b)}  Cooling for different atom positions, which corresponds to different polarizations. Best cooling occurs at linear light,  and give $\langle n \rangle\approx 1$.    \textbf{(c)} Clebsch-Gordon coefficients for $J=1/2$ to $J'=3/2$. } 
  \label{fig:pg1o2}
\end{figure}

In Fig.~\ref{fig:pg1o2}, we also investigate the case of  $J=1/2$ to $J'=3/2$ for lin$\perp$lin, which was studied in Ref.~\cite{cirac1993laser}  {and experimentally in ions in Refs.~\cite{joshi2020polarization, li2022robust}}.  There they found a lower limit of $\langle n \rangle \approx 1$.   {The picture they give is shown in (a), where the two spin states are displaced and a position dependent optical pumping gradually lowers the energy.}   For $J=1/2$, there is no cooling for $\spsm$, which requires at least three ground spins due to the tensor shift.   Because the polarization in lin$\perp$lin changes with position, we plot in (b) the cooling as a function of atom position, with $\phi = 0$ linear and $\phi = \pi/2$ circular. Similar to the other PG model, the population converges to $\langle n \rangle \approx 1$.  Interestingly though, this occurs for when the polarization is linear, not circular.  This is because circular polarization cooling also requires three spins because the Raman transition supplies two angular momenta.  The cooling still occurs for when the light shift is equal to the trapping frequency~\cite{cirac1993laser}. However, we do find that for larger spin states, circular polarization is the best cooling, in agreement with our model. 

%%%%%%%%%%%%%%%%%%%%%%%%%%%%%%%%%%%%%%%%%%%%%%
\subsection{ \texorpdfstring{$\Lambda$}{Lambda}-GM and EIT Cooling }  \label{subsec:lambda}
\begin{figure*}[tb]
  \centering
  \includegraphics[width=0.99\linewidth]{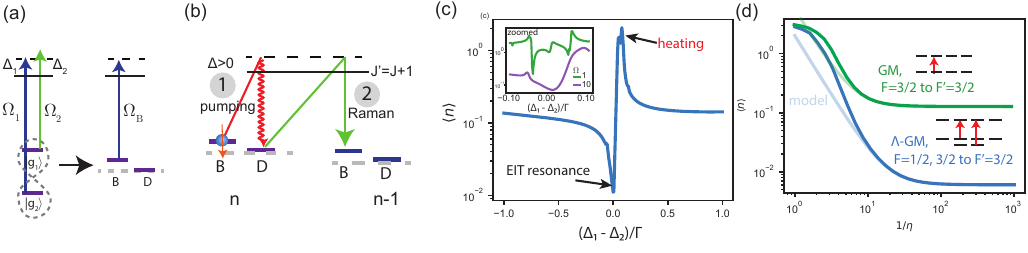}
  \caption{  {\textbf{(a)} EIT and $\Lambda$-enhanced GM decrease temperature by creating dark states from superpositions of ground states using two coherent lasers.  \textbf{(b)} EIT cooling uses two lasers with Rabi frequencies $\Omega_1$ and $\Omega_2$ to address the two ground states, driving a two-photon transition from $|B, n \rangle $ to $|D, n-1 \rangle $. }   \textbf{(c)} Master equation simulations of population versus the two-photon detuning. Best cooling occurs at the two-photon resonance when $\Delta_1 = \Delta_2$.  A heating resonance occurs to the right.    Parameters are $\eta = 0.05$, $\Delta = 10 \Gamma$, $\nu = 0.05 \, \Gamma$, $\Omega_1 = \Omega_2$, and 8 harmonic states.  The cooling power is set so that the light shift equals the trapping frequency.  The inset shows that as the cooling power is lowered, the resonance shifts over. When the light shift becomes small compared to the trapping frequency, the two-photon detuning must compensate to drive $n \rightarrow n-1$.  \textbf{(d)}  Comparison of GM and $\Lambda$-GM. The $\Lambda$-GM model contains a $F=1/2$ and $F=3/2$ ground state with two laser frequencies.  The GM model just has a single ground state $F=3/2$.   Light dashed is simple model from Eq.~\ref{eq:simplemodel} with effective parameters.} 
  \label{fig:EIT}
\end{figure*}

 {The spin models reveal two strategies for lowering the temperature: first, to make the lower energy state dark to optical pumping back into the other spin state, and second to decrease the recoil heating from spontaneous emission.   In terms of the collapse operator, these correspond to the position-independent and position-dependent terms.} Gray molasses (GM) cooling generates dark states through optical angular momentum selection rules, for example the $m_F=0$ to $m_F'=0$ state is dark for $F'=F$.  {Electromagnetically induced transparency (EIT) cooling and $\Lambda$-type gray molasses ($\Lambda$-GM) create dark states by creating superpositions of multiple ground states in an EIT configuration that are dark to the scattering.  The second set of ground states are addressed with a second coherent laser, as illustrated in Fig.~\ref{fig:EIT}(a). } These dark states can be viewed as arising from the interference of the two lasers coupling the ground states to a common excited state.

Historically, the term ``EIT cooling" has been primarily associated with trapped ions~\cite{roos2000experimental, Lin2013Sympathetic}, where the two ground states are typically different Zeeman sublevels. In contrast, ``$\Lambda$-GM" is commonly used in the context of neutral atoms~\cite{Brown2019Gray, huang2022gray, blodgett2023imaging}, where the two ground states belong to different hyperfine manifolds of the ground state.  {$\Lambda$-GM modifies conventional GM with a second coherent laser to address the different hyperfine states, creating dark states with multiple spin states.  EIT cooling often instead refers to optical pumping to a single spin state and only using two ground spin states. 
}

Cooling to the motional ground state using EIT has been theoretically predicted and achieved experimentally in $\text{Ca}$ ion using Zeeman sublevels of the D1 line~\cite{Giovanna2000Ground, roos2000experimental}. $\Lambda$-enhanced GM cooling on the other hand is proven effective not only for bound atoms \cite{blodgett2023imaging} but also for cooling and imaging molecules in optical traps~\cite{cheuk2018lambda}.

 {To model $\Lambda$-GM or EIT cooling, we use the standard EIT formalism, for example in Ref.~\cite{steck}, to transform the system to the bright and dark basis.  Then we apply the formalism from this paper to show that it produces similar position dependent and independent operators as in the spin cooling model.  We will then discuss the fundamental limits of this type of cooling.}

EIT addresses two ground states $|g_1 \rangle$ and $|g_2 \rangle$ with two different Rabi frequencies $\Omega_1$ and $\Omega_2$, as shown in Fig.~\ref{fig:EIT}(a).  These two states are typically from different hyperfine ground states.  Each ground state is addressed by a laser frequency to a single excited state with corresponding detunings $\Delta_1$ and $\Delta_2$. The Hamiltonian is written with $g_1$ in the rotating frame of laser one and $g_2$ in the rotating frame of laser two,
\begin{equation}
H_0 = \nu a^\dag a + \Delta_1 |g_1 \rangle \langle g_1 | + \Delta_2 |g_2 \rangle \langle g_2 |.
\end{equation}
We have two dipole lowering operators, $\hat{D}_1 = |g_1 \rangle \langle e|$ and $\hat{D}_2 = |g_2 \rangle \langle e|$. The interaction Hamiltonian is
\begin{equation}
H_{I} = \frac{\Omega_1}{2} (\hat{D}_1 ^\dag e^{i k_1 \hat{x}} + {h.c.} ) + \frac{\Omega_2}{2} (\hat{D}_2 ^\dag e^{i k_2 \hat{x}} + {h.c.} ).
\end{equation}
The two collapse operators represent the spontaneous emission of the single excited state to both ground states,
\begin{equation}
L_{1, \pm} = \sqrt{\Gamma} \,  e^{\pm ik_1\hat{x}} \hat{D}_1,
\quad L_{2, \pm} = \sqrt{\Gamma}  \, e^{\pm ik_2\hat{x}} \hat{D}_2.
\end{equation}

Next we transform the ground states into the bright and dark basis. We perform a unitary transformation of the two ground states into two superpositions so that only one of them is coupled to the two lasers, and the other is not.  The superposition that is dark to the light is the EIT dark state, shown in Fig.~\ref{fig:EIT}(a). The dark state is dark due to the destructive interference of the population in the excited state from the two ground states. 

The basis transformation to the bright and dark basis is 
\begin{align}
|g_{\scriptscriptstyle B}\rangle &= \frac{1}{\Omega_\text{rms}} \left(\Omega_1 \, |g_1 \rangle + \Omega_2 \, |g_2 \rangle \right) \nonumber   \\
|g_{\scriptscriptstyle D}\rangle &=\frac{1}{\Omega_\text{rms}} \left(-\Omega_2 \, |g_1 \rangle + \Omega_1 \, |g_2 \rangle \right) \\
|e \rangle &= |e \rangle \nonumber
\end{align}
where $\Omega_\text{rms} = \sqrt{ \Omega_1^2 + \Omega_2^2}$.  
Under this unitary transformation, the bare Hamiltonian becomes
\begin{gather} 
\tilde{H}_0 = \nu a^\dag a + \Delta_{\scriptscriptstyle B} |g_{\scriptscriptstyle B} \rangle \langle g_{\scriptscriptstyle B} | + \Delta_{\scriptscriptstyle D} |g_{\scriptscriptstyle D} \rangle \langle g_{\scriptscriptstyle D} |  \nonumber \\
+\Omega_{\scriptscriptstyle BD} \bigl( | g_{\scriptscriptstyle B} \rangle \langle g_{\scriptscriptstyle D}| +| g_{\scriptscriptstyle D} \rangle \langle g_{\scriptscriptstyle B}| \bigr).   \label{eq:H0tilde2}
\end{gather}
The energies of the bright and dark states are
\begin{gather}
\Delta_{\scriptscriptstyle B} =\left( \Omega_1^2 \,\, \Delta_1 +  \Omega_2^2 \,\, \Delta_2 \right)/\Omega_\text{rms}^2 \nonumber \\
\Delta_{\scriptscriptstyle D} = \left( \Omega_2^2 \,\,\Delta_1 +  \Omega_1^2\,\, \Delta_2 \right)/ \Omega_\text{rms}^2,
\end{gather}
and the coupling between the dark and bright states is 
\begin{equation}
    \Omega_{\scriptscriptstyle BD} = (\Delta_2 - \Delta_1) \frac{\Omega_1 \Omega_2}{\Omega_1^2 + \Omega_2^2}.
\end{equation}
When $\Delta_1 = \Delta_2$, both states are detuned by the same amount from the excited state and the Raman transition is resonant.  Then the coupling in the bare Hamiltonian is $\Omega_{BD} = 0$.

The transformed atom-field interaction is 
\begin{gather}
   \tilde{H}_I =\frac{1}{2 \Omega_\text{rms} } \left[  \hat{D}_{B}^\dag \, \left( \Omega_1^2 \, e^{i k_1 \hat{x}}  + \Omega_2^2 \, e^{i k_2 \hat{x}}   \right)   \right. \nonumber \\
   \left. +  \hat{D}_{D}^\dag  \Omega_1 \Omega_2 \, \left(   e^{i k_2 \hat{x}}  -  e^{i k_1 \hat{x}}   \right)      + h.c.  \right].
\end{gather}

% \begin{gather}
%     \tilde{H}_I = \Omega \left(  \cos(k\hat{x}) D_{B}^\dag   +  \sin(k \hat{x}) D_{D}^\dag  \right) \\
%     \approx \Omega \left(  D_{B}^\dag   + k \hat{x} \,\, D_{D}^\dag  \right) .
% \end{gather}
% \begin{gather}
%     \tilde{H}_I = \frac{\Omega_\text{rms}}{2} \hat{D}_B^\dag  \\
%     + i x \left[  \hat{D}_B^\dag \frac{(k_1 \Omega_1^2 + k_2 \Omega_2^2)}{2 \Omega_\text{rms}} + \hat{D}_D^\dag \frac{\Omega_1 \Omega_2 (k_1 - k_2 ) }{2 \Omega_\text{rms}}   \right]   + h.c.
% \end{gather}
%For co-propagating light, $k_1 = k_2$ and then the dark state is not coupled to any other state. Cooling does not occur. However, for counter-propagating light $k_2 = - k_1$, the dark state is coupled. If we assume counter-propagating light and  equal Rabi frequencies $\Omega_1 = \Omega_2$, then this simplifies to 
If we assume that the two lasers  are counter-propagating $-k_1 = k_2=k$, assume the Rabi frequencies are the same, and expand the interaction Hamiltonian to first order in the LD parameter, we get
\begin{gather}
    \tilde{H}_I = \frac{\Omega}{\sqrt{2}} \left( \hat{D}_B^\dag  
    +i k x \hat{D}_D^\dag    + h.c. \right)
\end{gather}
 { From this form, we can draw some interesting conclusions.  If the atom is not oscillating, only the bright state is coupled to the excited state.  The dark is state is completely dark and population will quickly optically pump to it.  
When the atom oscillates, the dark state is driven to one of the sidebands.}

% An interesting case is if, instead, the lasers are co-propagating with $k_1 = k_2$,  then $H_I = \Omega  D_B^\dag e^{ikx}$, and the dark state is dark even to $n$ changing transitions.  Co-propagating light cannot change the momentum, prohibiting cooling. 

Next, we can adiabatically eliminate the excited state using Eq.~\ref{eq:Hproj} and \ref{eq:Lproj} to get the effective ground state dynamics.
In the bright and dark basis, the effective ground state Hamiltonian is
\begin{gather} 
    \tilde{H}_\text{eff} =\nu a^\dag a + \frac{\Omega_1^2 + \Omega_2^2}{2\Delta } 
     |g_{\scriptscriptstyle B}\rangle \langle g_{\scriptscriptstyle B}| \nonumber \\ 
     + \left( i k \hat{x} \, \Omega_{r} \,  |g_{\scriptscriptstyle B} \rangle \langle g_{\scriptscriptstyle D}|  + h.c. \right)  
\end{gather} 
and the collapse operators are
\begin{gather}
    \tilde{L}_{\text{eff}, \pm ,i}  =  \sqrt{\gamma_r}\,  \hat{D}_{i} \left( \, \hat{D}^\dag_{B}  \pm ik\hat{x} (\, \hat{D}_{B}^\dag + \frac{ 2 \Omega_1 \Omega_2}{\Omega^2_\text{rms}} \, \hat{D}^\dag_{D}  )\right)  ,
\end{gather}
where $i=1,2$ are the two ground states, and the off-resonant scattering rate from both beams is
\begin{equation}
    \gamma_{r} =  \Gamma\,  \frac{ \Omega_1^2 + \Omega_2^2 }{ \Delta^2},
\end{equation}
and the Raman Rabi rate between the bright and dark state is 
\begin{equation}
    \Omega_{r} = \frac{\Omega_1 \Omega_2}{2\Delta}
\end{equation} 

 { The form of the operators are similar to the GM spin cooling model. In the position independent effective Hamiltonian, the dark state is not driven and has no light shift.  That makes it similar to the GM case, requiring blue-detuned light to make the bright state higher in energy.   Similarly, the position independent collapse operators only couple the bright state, with an off-resonant scattering rate  for $\Delta \gg \Gamma$.  However the position dependent term couples the bright and dark state to different motional states, resulting in recoil heating at a rate $\approx \gamma_s \eta^2$.  The effective Raman transition between $|g_D, n \rangle$ and $|g_B, n-1 \rangle$ is $\Omega_r$.    Just as we found in the spin cooling section, the transfer rate comes from the coherent coupling and the optical pumping, resulting in a transfer rate of $\Gamma_r \approx \Omega_r^2/ \gamma_r$.  }

 {The picture for cooling is in Fig.~\ref{fig:EIT}(b). Population is pumped to the dark state.  Then the two-photon process couples $|D, n \rangle$ to $|B, n-1 \rangle$.  
}

 {
In Fig.~\ref{fig:EIT}(c), we simulate the cooling performance as a function of the two-photon detuning.  These simulations are for the full ground and excited master equation without making the LD approximation. In the main plot, the bright state light shift is tuned to be equal to the trapping frequency, in which case optimal cooling occurs when $\Delta_1 = \Delta_2$. The cooling has an EIT resonance at $\Delta_1 = \Delta_2$.  The linewidth of this resonance is given by the off-resonant scattering rate $\gamma_s$.  There is a heating resonance nearby when $\Delta_1 - \Delta_2 =2 \nu $.  
}

The inset shows in Fig.~\ref{fig:EIT}(c) simulates the cooling performance as the cooling light power is decreased. When the light shift is not equal to the trapping frequency, the two-photon detuning can be adjusted to bring the system back into resonance and recover the cooling efficiency.

Fig.~\ref{fig:EIT}(d) compares the optimal cooling performance for the $F=1/2, 3/2$ to $F'=3/2$ system with and without the $\Lambda$-enhancement (i.e., with and without the $F=1/2$ state). The inclusion of the $F=1/2$ state in the $\Lambda$-configuration significantly improves the cooling performance. The simulation results are also fitted to the theoretical model to extract the effective cooling parameters.  The light dashed lines are fitting the simple spin model from Eq.~\ref{eq:simplemodel} with parameters $\gamma_h/\gamma_+ =2.0$ and $\gamma_-/\gamma_+ = 0.006$ for $\Lambda$-GM, and $\gamma_s/\gamma_+ = 5.0$ and $\gamma_-/\gamma_+ = 0.11$ for GM.

The effects of EIT on the cooling performance can now be clearly seen. By creating darker reservoir states, EIT decreases the effective asymmetry between the pumping rates, $\gamma_{-}/\gamma_{+}$. As a result, the fundamental limit for EIT cooling is similar to that of GM cooling, but reduced by the factor by which the dark states are darker. The cooling efficiency still depends on the inverse square of the Lamb-Dicke parameter, $1/\eta^2$. In Fig.~\ref{fig:EIT}(d), we plot the cooling performance for the $F=1/2, 3/2$ to $F'= 3/2$ system, comparing the cases with and without the $F=1/2$ state. The inclusion of the $F=1/2$ state in the $\Lambda$-configuration leads to a significant improvement in the cooling efficiency, demonstrating the power of EIT in enhancing the performance of sub-Doppler cooling techniques.

% XX1XX
 {
One important requirement for EIT cooling is that the two lasers are coherent and at the same frequency. A non-zero two-photon detuning contributes to a bright-dark coupling in Eq.~\ref{eq:H0tilde2} with a rate \( \Omega_{BM} \approx \Delta_1 - \Delta_2 \). As the phase between the two lasers slips, the bright and dark basis changes, also producing this coupling. The other bright-dark coupling in the effective Hamiltonian scales as \( \eta \Omega_r \). For resonance \( \Omega_r \approx \nu \), this means that \( \Delta_1 - \Delta_2 \) must be less than \( \eta \nu \) to be negligible. For typical trapping frequencies of 100 kHz, this implies that the lasers must be coherent to within \( \sim 1 \) kHz, which occurs naturally if the the two frequencies are generated from the same light using an EOM or AOM.
}

%%%%%%%%%%%%%%%%%%%%%%%%%%%%%%%%%%%%%%%%
\subsection{Raman-sideband cooling}  \label{subsec:raman}
\begin{figure}[b]
  \centering
  \includegraphics[width=0.8\linewidth]{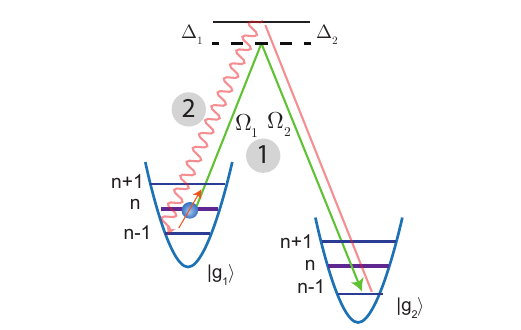}
  \caption{ Raman-sideband cooling. The first step is a Raman transition between spin states and decreasing the motional state by 1. The second step is spontaneous emission back to original spin, which keeps atom in $n-1$ in the Lamb-Dicke regime.  The cycle repeats until the population is in the ground state.  } 
  \label{fig:RSB}
\end{figure}   

Raman sideband (RSB) cooling has high fidelity ground state preparation in neutral atoms~\cite{Hamann1998Resolved, kerman2000beyond, kerman2002raman}, ions~\cite{monroe1995resolved}, optical tweezers~\cite{kaufman2012cooling, thompson2013coherence, yu2018motional}, and molecules~\cite{bao2023raman, Lu2024Raman}. The working principle of RSB cooling is illustrated in Fig.~\ref{fig:RSB}(a).

In RSB cooling, two coherent, far-detuned lasers drive the transition from $|g_1, n \rangle$ to $|g_2, n-1\rangle$, where $|g_1\rangle$ and $|g_2\rangle$ are two ground states, and $n$ represents the motional quantum number. The light shifts and scattering rates from these cooling lasers are typically negligible due to the large detuning. A third laser, tuned closer to resonance, performs optical pumping and exploits selection rules to ensure that the dark state remains dark.

For the RSB case, we instead derive an effective ground state basis in the original $g_1$, $g_2$ basis instead of the bright and dark basis. 
In the original basis, the effective ground state Hamiltonian is
\begin{gather}  \label{eq:EITeffective}
    H_\text{eff} = \nu \hat{a}^\dag \hat{a} + (\Delta_1 -\frac{\Omega_1^2}{\Delta}) |g_1 \rangle \langle g_1| +(\Delta_2 - \frac{\Omega_2^2}{\Delta} )|g_2 \rangle \langle g_2|  \nonumber \\
     -\frac{\Omega_1 \Omega_2}{\Delta} \left( |g_1 \rangle \langle g_2| e^{i(k_2 - k_1) \hat{x}}  + h.c. \right)  .
\end{gather}
The light shifts $\Omega_1^2/\Delta$ and $\Omega_2^2/\Delta$ shift the two states.  
The effective Raman coupling strength between the two ground states is given by $\Omega_R = \Omega_1 \Omega_2/ \Delta$, where $\Omega_1$ and $\Omega_2$ are the Rabi frequencies of the two cooling lasers, and $\Delta$ is the detuning from the excited state. 

The transition matrix elements between the two ground states are
\begin{gather}
\langle g_1, n_1 | H_\text{eff} | g_2, n_2 \rangle =
-\frac{\Omega_1 \Omega_2}{\Delta} \langle n_1| e^{i(k_1 - k_2) \hat{x}} | n_2 \rangle,
\end{gather}
where $k_1$ and $k_2$ are the wave vectors of the two cooling lasers. By carefully tuning the laser frequencies and polarizations, this Raman coupling can be engineered to drive transitions that reduce the motional quantum number, leading to cooling.

In 3D, the Raman coupling term becomes $e^{i (\bm{k}_1 -\bm{k}_2)\cdot \hat{\bm{r}} } \approx 1 + i \hat{x} \, (\bm{k}_1 -\bm{k}_2) \cdot  \hat{\bm{x}}$, which is related to the projection of the difference wave-vector onto the cooling axis. Multiple laser beams are typically used so that pairs of beams have projections on all three axes, enabling 3D cooling.

RSB cooling improves upon EIT and $\Lambda$-GM cooling techniques, which introduced dark states in an electromagnetically induced transparency (EIT) configuration. While EIT and $\Lambda$-GM achieved lower temperatures, they were limited by spontaneous emission from the dark state, parameterized by the ratio $\eta^2 \gamma_h/\gamma_+$, where $\eta$ is the Lamb-Dicke parameter, $\gamma_h$ is the heating rate, and $\gamma_+$ is the pumping rate.

RSB cooling addresses this limitation by increasing the detuning of the cooling lasers until the collapse operator $L$ becomes negligible, effectively eliminating spontaneous emission and repumping from the cooling lasers. The additional repump beam, optimized using angular momentum selection rules, ensures that the state $|g_1 \rangle$ remains dark to it, making the dark state immune to position-dependent spontaneous emission.  {PG, GM, and $\Lambda$ had a position dependent collapse operator that drove heating in the dark state.  This is because they use counter-propagating light that coherently couple spins to also optically pump.  RSB uses a separate optical pumping beam that is completely dark to the dark state.  }

An effective spin model for RSB cooling can be created, where the light shift from the cooling light is ignored, and the two-photon detuning is adjusted to bring the $|g_1, n \rangle$ to $|g_2, n-1 \rangle$ transition into resonance:
\begin{gather}      
    H = \nu \hat{a}^\dag \hat{a} + (\Delta_1 - \Delta_2) |g_2 \rangle \langle g_2| + \Omega_R  k \hat{x} \,  \hat{F}_x   \nonumber \\
    L_{+} =\sqrt{\gamma_{\scriptscriptstyle +}}\,\, \hat{F}_{+}( 1 + i k \hat{x}). 
\end{gather}

While RSB cooling offers significant advantages, its fundamental limitations are often technical rather than theoretical. The coherence time between the two ground states is typically limited to a few milliseconds due to factors such as magnetic field fluctuations or laser decoherence. For efficient cooling, the Raman Rabi frequency $\Omega_R$ should exceed 10 kHz. However, the Rabi frequency for the $n$ to $n$ transition is larger than that of the $n$ to $n-1$ transition by a factor of $1/\eta$, resulting in both transitions being driven simultaneously. Further analysis is needed to determine the cooling limit in the presence of these technical constraints.

While RSB cooling offers a pathway to overcome the limitations of EIT and $\Lambda$-GM cooling schemes, its implementation is more complex experimentally. However, it potentially enables more efficient ground state cooling in a wider range of atomic and molecular systems.

 {Similar to $\Lambda$-GM, RSB also requires phase coherence of the laser.  It also requires that $\Omega_r$ is less than the trapping frequency.  However, RSB is usually performed with a separate set of lasers than the MOT beams. 
}

%\note{discussion of n to n transition}

%%%%%%%%%%%%%%%%%%%%%%%%%%%%%%%%%

\section{Analysis}
 {
In this paper, we have demonstrated that polarization gradient (PG), gray molasses (GM), $\Lambda$-enhanced gray molasses ($\Lambda$-GM), electromagnetically induced transparency (EIT), and Raman sideband (RSB) cooling can all be understood within a unified framework of spin cooling. Our formalism reveals that all these two-photon cooling schemes share a fundamental similarity: they involve a Raman transition between neighboring motional states with different spin states.

After adiabatically eliminating the excited state and expanding to first order in the Lamb-Dicke (LD) parameter, each cooling model exhibits a similar structure, featuring a Hamiltonian and collapse operator with both position-independent and position-dependent terms. The key ingredients for effective cooling are: (1) a differential light shift between spin states equal to the trapping frequency, ensuring resonance between $|g_2,n\rangle$ and $|g_1, n-1\rangle$, (2) preferential optical pumping ($\gamma_-/\gamma_+ < 1$) to the lower energy state, and (3) coherent coupling between the resonant states.

We have identified three primary factors that determine the final temperature: a large asymmetry in optical pumping, a small LD parameter, and maximizing the darkness of the lower energy state. Different cooling techniques implement these principles in various ways, with close-to-resonance cooling schemes like PG and GM utilizing light shifts to bring the states into resonance, while further detuned methods such as RSB cooling use the two-photon detuning of two frequencies and two far-detuned ground states.

PG and GM are experimentally efficient, using a pair of counter-propagating beams for all three requirements. However, PG cooling is fundamentally limited by unity pumping asymmetry, with $\langle n \rangle \approx \gamma_-/ \gamma_+ \approx 1$ for $F$ to $F+1$ transitions. GM differs in the darkness of the lower energy state, utilizing natural dark states of the ground state manifold when $F' \le F$. In the limit of $\gamma_- \ll \gamma_+$, GM's fundamental limit is $\langle n \rangle \approx \eta^2 \gamma_h/\gamma_+$, where $\gamma_h / \gamma_+$ is the ratio of the effective scattering rate of the dark state to the pumping rate, typically close to unity for GM. For heavy atoms with $\eta < 0.1$, this can lead to ground state preparation.

$\Lambda$-GM maintains the same scaling as GM but enhances cooling by introducing more dark states and a second coherent laser to address the other hyperfine ground state. The relative detuning in these EIT configurations in principle relaxes the condition that light shifts must equal the trapping frequency (the case for GM), as the relative detuning of the two lasers can compensate. 

However, these schemes still rely on these same two counter-propagating lasers for optical pumping. Despite the improvements offered by GM and $\Lambda$-GM, spontaneous emission in these schemes still drives the population out of the dark state through either spontaneous emission from the first-order LD parameter or, in the effective ground state formalism, the position-dependent component of the collapse operator. The dark $n=0$ state is thus not completely dark, as it can still be driven through the $n=1$ excited state.

RSB cooling represents the ultimate limit of cooling, detuning so far from the excited state that the spontaneous emission of the Raman coupling light is negligible. It separates the cooling process into three distinct steps: (1) the relative detuning of two beams brings the trapping frequencies of two spin states into resonance, (2) these beams create coherent transfer, and (3) an additional resonant laser provides optical pumping, specifically designed to maximize the darkness of the dark state. The re-pumping is achieved through a separate spontaneous emission beam optimized to make the dark state truly dark.

In practical laboratory settings, PG, GM, and $\Lambda$-GM are more convenient due to their operational simplicity and similarity to the magneto-optical trap (MOT) configuration. However, RSB is still employed for high ground state preparation despite its significant experimental overhead. Our work demonstrates that gray molasses cooling combined with the $\Lambda$-enhancement can achieve significant ground state populations, raising the question of how schemes resembling MOT configurations can be adapted for near-ground state cooling.

One potential strategy would be to use GM or $\Lambda$-GM, detuned far enough to make the scattering rate negligible, then add a resonant beam for spontaneous emission.  The goal is to eliminate the position-dependent collapse operator that drives the dark transition, which is the only coupling out of the dark ground state.  This perhaps could be accomplished by having separate optical pumping beams that are co-propagating and drive only the bright state.   In the simple model picture,  this would effectively create a collapse operator $L_+ = \sqrt{\gamma_+} e^{ik\hat{x}} \hat{F}_+$ similar to the RSB model, where the second term doesn't drive the dark state. 

This approach could potentially combine the advantages of both worlds, offering the experimental simplicity of GM and $\Lambda$-GM with the improved ground state preparation of RSB. Further investigation of this possibility will be the subject of future experimental and theoretical work, as we continue to explore how these cooling schemes can be optimized and combined to achieve even lower temperatures and higher ground state populations in trapped atomic systems.}

\section{Conclusion}  \label{sec:conclusion}
This study presents a comprehensive theoretical framework for major cooling mechanisms in neutral atom tweezers, uncovering shared principles across diverse techniques. Our approach combines detailed full-level structure simulations with a simplified spin model, offering novel insights for optimizing cooling schemes. By extending previous research to encompass arbitrary spins and polarizations, we show that gray molasses cooling may potentially reach the ground state, challenging conventional limits and opening up new avenues for exploration.

Future research will investigate polarization gradients in three-dimensional beam configurations and examine deviations from the idealized one-dimensional cases presented here. We also plan to conduct simulations for specific alkali atoms, including all relevant ground and excited states, to assess the impact of multiple excited states and cross-coupling between lasers addressing different hyperfine ground states. These simulations will help establish fundamental cooling limits for each atomic species for various techniques. Future research will explore innovative cooling approaches that blur the boundaries between established techniques.

Rapid, robust, and efficient ground state optical cooling will increase the coherence and fidelity of experiments throughout the ultracold field. Reducing thermal motion in Rydberg atoms will pave the way for higher-fidelity quantum gates. Rapid optical cooling techniques could also reduce the long times associated with thermal evaporation for producing degenerate gases. 

\section*{Acknowledgements}
This work was supported by the NSF Career Award (Award No. 0543784). 
We express our gratitude to Yichao Yu for valuable early discussions regarding the relationship between spin cooling and Raman-sideband cooling.

\appendix*
\input{appendix}

\bibliographystyle{apsrev4-2}
\bibliography{references_master}

\end{document}

%% file: appendix.tex
\section*{Appendix A}
\subsection*{Dipole spin operator}
For the atom spin, we will use the hyperfine spin states $\bm{F} = \bm{J} + \bm{I}$. Note that setting $I=0$ recovers the $J$ formalism for comparison. The excited states are $|F_e, m_e \rangle$ and ground states are $|F_g, m_g \rangle$. 
 The dipole operator in Eq.~\ref{eq:dipoleoperator} is in this spin basis and using the Wigner-Eckart theorem is 
\begin{equation} \label{eq:Dq}
\hat{D}_{q} = \sum_{m_g m_e} O^{J_g J_e}_{I F_g F_e} \,\, \langle F_g m_g|F_e m_e;1q\rangle \,\, |F_g m_g\rangle\langle F_e m_e|.
\end{equation}
The oscillator strength coefficients are given by
\begin{equation}
\setlength{\arraycolsep}{2pt} % Reduces spacing between columns
\renewcommand{\arraystretch}{0.8} % Reduces vertical stretching of rows
O^{J_g J_e}_{ I F_g F_e}  =  (-1)^{F_e \!+ \!J_g + \!I \! + \! 1}  \sqrt{(2F_e \!+ \! 1)(2 \! J_g \! + \! 1)}
\begin{Bmatrix}
J_e & J_g & 1 \\
F_g & F_e & I  
\end{Bmatrix}  , 
\end{equation}
which are on the order of unity and give the strength of a $F_g$ to $F_e$ transition relative to the $J_g$ to $J_e$ transition with $\sum_{F_e} |O^{J_g J_e}_{I F_g F_e}|^2 = 1$. For more details, see Refs.~\cite{deutsch2010quantum} and \cite{steck}. 

%The dipole operators expressed in terms of the spherical vectors are  $\hat{D}_z = \hat{D}_0$ and  $\hat{D}_{x,y} = \frac{1}{\sqrt{2}}(\hat{D}_{1} \pm \hat{D}_{ -1} )  $

\subsection*{Spin tensor coefficients}
The coefficients are determined by the Wigner-Eckart reduced matrix in the ground and excited spins.  For the case of $J,F$ in the ground and $J',F'$ in the excited, the coefficients are~\cite{deutsch2010quantum}:
\begin{gather} 
C^{(0)} = (-1)^{3F-F'+1} \sqrt{\frac{1}{3}} \frac{2F' + 1}{\sqrt{2F + 1}} \\
\times
\begin{pmatrix}
F & 1 & F' \\
1 & F & 0
\end{pmatrix} 
\left| \begin{Bmatrix}
1 & F & F' \\
0 & 1/2 & F
\end{Bmatrix} \right|^2, \nonumber \\
C^{(1)} = (-1)^{3F-F'} \sqrt{\frac{3}{2}} \frac{2F' + 1}{\sqrt{F(F + 1)(2F + 1)}} 
\\
\times \begin{pmatrix}
F & 1 & F' \\
1 & F & 1
\end{pmatrix} 
\left| \begin{Bmatrix}
1 & F & F' \\
0 & 1/2 & F
\end{Bmatrix} \right|^2, \nonumber \\
C^{(2)} = (-1)^{3F-F'}
\sqrt{\frac{30}{F(F + 1)(2F + 1)(2F - 1)(2F + 3)}}    \\
\times(2F' + 1) \times
\begin{pmatrix}
F & 1 & F' \\
1 & F & 2
\end{pmatrix} 
\left| \begin{Bmatrix}
1 & F & F' \\
0 & 1/2 & F
\end{Bmatrix} \right|^2. \nonumber
\end{gather}